\documentclass[a4paper]{article}

\usepackage[top=1in, bottom=1.25in, left=1.25in, right=1.25in]{geometry} 
\usepackage{graphicx}
\usepackage{tikz} 
\usetikzlibrary{positioning, chains, calc, arrows, decorations.pathreplacing, fit, patterns, graphs}
\usepackage{amssymb}
\usepackage{amsmath}
\usepackage{amsthm}
\usepackage{tabularx}
\usepackage{paralist}
\usepackage{enumerate}
\usepackage{microtype}

\usepackage{url}
\usepackage[hidelinks]{hyperref}                
\usepackage{authblk}

\DeclareMathOperator{\npclass}{\mathsf{NP}}
\DeclareMathOperator{\pspaceclass}{\mathsf{PSPACE}}
\DeclareMathOperator{\pclass}{\mathsf{P}}
\DeclareMathOperator{\wclass}{\mathsf{W}}

\DeclareMathOperator{\pat}{\textsf{Pat}}
\DeclareMathOperator{\loc}{\textsf{loc}}
\DeclareMathOperator{\noncross}{\textsf{nc}}
\DeclareMathOperator{\regular}{\textsf{reg}}
\DeclareMathOperator{\reg}{\textsf{reg}}
\DeclareMathOperator{\stronglynested}{\textsf{snest}}
\DeclareMathOperator{\nested}{\textsf{nest}}
\DeclareMathOperator{\repeatedvar}{\textsf{repv}}
\DeclareMathOperator{\scd}{\textsf{scd}}
\DeclareMathOperator{\mildlyentwined}{\textsf{ment}}
\DeclareMathOperator{\outp}{\textsf{outp}}

\DeclareMathOperator{\allpatterns}{\textsf{Pat}}

\DeclareMathOperator{\opt}{\textsf{opt}}

\DeclareMathOperator{\pathwidth}{\textsf{pw}}
\DeclareMathOperator{\cutwidth}{\textsf{cw}}

\newcommand{\locProb}{\textsc{Loc}}
\newcommand{\minlocProb}{\textsc{MinLoc}}

\newcommand{\cutwidthProb}{\textsc{Cutwidth}}
\newcommand{\pathwidthProb}{\textsc{Pathwidth}}
\newcommand{\minPathwidthProb}{\textsc{MinPathwidth}}
\newcommand{\minCutwidthProb}{\textsc{MinCutwidth}}

\DeclareMathOperator{\match}{\textsc{Match}}

\DeclareMathOperator{\scope}{sc}

\DeclareMathOperator{\var}{\textsf{var}}
\DeclareMathOperator{\alphabet}{\textsf{alph}}

\DeclareMathOperator{\bigO}{O}

\DeclareMathOperator{\equality}{\mathsf{equ}}
\DeclareMathOperator{\neighbour}{\mathsf{nei}}
\DeclareMathOperator{\pattern}{\mathsf{pat}}
\DeclareMathOperator{\word}{\mathsf{wo}}

\newcommand{\ta}{\ensuremath{\mathtt{a}}}
\newcommand{\tb}{\ensuremath{\mathtt{b}}}
\newcommand{\tc}{\ensuremath{\mathtt{c}}}

\newcommand{\tg}{\ensuremath{\mathtt{g}}}

\newtheorem{theorem}{Theorem}

\theoremstyle{remark}
\newtheorem{example}[theorem]{Example}
\newtheorem{remark}[theorem]{Remark}
\newtheorem{problem}[theorem]{Problem}

\title{Matching Patterns with Variables}

\author[1]{Florin~Manea}

\author[2]{Markus L. Schmid}

\affil[1]{Kiel University, Kiel, Germany, \texttt{flm@informatik.uni-kiel.de}}
\affil[2]{Trier University, Trier, Germany, \texttt{MLSchmid@MLSchmid.de}}

\date{\vspace{-1cm}}

\begin{document}

\maketitle

%\title{Matching Patterns with Variables}
%
%\author{Florin Manea\inst{1}\orcidID{0000-0001-6094-3324} \and
%Markus L. Schmid\inst{2}\orcidID{0000-0001-5137-1504}}
%
%\authorrunning{F. Manea and M. L. Schmid}
%
%\institute{Kiel University, Germany, \email{flmanea@gmail.com} \and
%Trier University, Germany, \email{MLSchmid@MLSchmid.de}}
%
%\maketitle              % typeset the header of the contribution

\begin{abstract}
A pattern $\alpha$ (i.\,e., a string of variables and terminals) matches a word $w$, if $w$ can be obtained by uniformly replacing the variables of $\alpha$ by terminal words. The respective matching problem, i.\,e., deciding whether or not a given pattern matches a given word, is generally $\npclass$-complete, but can be solved in polynomial-time for classes of patterns with restricted structure. In this paper we overview a series of recent results related to efficient matching for patterns with variables, as well as a series of extensions of this problem.

%\keywords{Combinatorial Pattern Matching  \and Patterns with Variables \and String Structural Parameters \and Efficient Algorithms \and NP-hardness}
\end{abstract}
\section{Introduction}
A \emph{pattern with variables}, called simply pattern in the context of this work, is a string that consists of \emph{terminal symbols} (e.\,g., $\ta, \tb, \tc$) and \emph{variables} (e.\,g., $x_1, x_2, x_3$). The terminal symbols are treated as constants, while the variables are to be uniformly replaced by strings over the set of terminals (i.\,e., different occurrences of the same variable are replaced by the same string); thus, a pattern is mapped to a terminal word. For example, $x_1 \ta \tb x_1 x_2 \tc x_2 x_1$ can be mapped to $\ta \tc \ta \tb \ta \tc \tc \ta \ta \tc \tc \ta \ta \ta \tc$ and $\tb \ta \tb \tb \ta \tc \ta \tb$ by the replacements $(x_1 \to \ta \tc, x_2 \to \tc \ta \ta)$ and $(x_1 \to \tb, x_2 \to \ta)$, respectively.

Patterns with variables appear in various areas of theoretical computer science, such as language theory (pattern languages~\cite{ang:fin2}), learning theory (inductive inference~\cite{ang:fin2,ng:dev,rei:dis,erl:lea}, PAC-learning~\cite{kea:apo}), combi\-natorics on words (word equations~\cite{kar:the,mat:fin}, unavoidable patterns~\cite{lot:alg:unPat}), pattern matching (generalised function matching~\cite{ami:gen,ord:apaJournal}), database theory (extended conjunctive regular path queries~\cite{bar:exp}), and we can also find them in practice in the form of extended regular expressions with backreferences~\cite{cam:afo,fri:mas,Fre2013}, used in programming languages like Perl, Java, Python, etc.\par

Generally, in all these contexts, patterns with variables are used to model various combinatorial pattern matching questions. For instance, searching for a word $w$ in a text $t$ can be expressed as testing whether the pattern $xwy$ can be mapped to $t$ and testing whether a word $w$ contains a cube is equivalent to testing whether the pattern $xy^3z$ can be mapped to $w$, such that $y$ is not mapped to an empty word. Not only problems of testing whether a given word contains a regularity or a motif of a certain form can be expressed by patterns, but also problems asking whether a word can be factorised in a specifically restricted manner can be modelled in this way. For instance, asking whether $x_1^2x_2^2\ldots x_k^2$ can be mapped to $w$, such that none of the variables $x_i$ are mapped to an empty word, is equivalent to asking whether the word $w$ can be factorised into $k$ non-empty squares. 

Unfortunately, deciding whether a given arbitrary pattern can be mapped to a given word, the \emph{matching problem}, is $\npclass$-complete \cite{ang:fin2}, whether we ask that the variables are mapped to non-empty words or not. This intractability result severely limits the practical application of patterns. Indeed, in many tasks related to applications of patterns, the matching problem is a necessary step, so the tasks become intractable as well. For instance, this is the case for the task of computing so-called descriptive patterns for finite sets of words (see~\cite{ang:fin2,FreydenbergerReidenbach2010,FreydenbergerReidenbach2013} for more information on descriptive patterns): one cannot solve this problem without solving a series of (general) pattern matching tasks~\cite{FeMaMeSc16_TCS}. A more detailed analysis of the complexity of the hardness of the matching problem will be presented in Section \ref{sec:hardness}. 

On the other hand, some strong restrictions on the structure of patterns yield subclasses for which the matching problem is tractable (i.e., can be solved in polynomial time). This is clearly the case of patterns where the number of different variables in the patterns is bounded by a constant, but more sophisticated and general such subclasses can be defined. We will discuss a series of results related to this topic in Sections \ref{sec:graphs}, \ref{sec:algos} and \ref{sec:local}. In our analysis, the most general class of patterns which allow for a polynomial-time pattern matching problem is defined by establishing a deep connection between strings/patterns and graphs, and considering only patterns which correspond to graphs with bounded structural parameters. As such, the subclass of {\em patterns with bounded treewidth}. The question of finding classes of patterns which can be matched in polynomial time but do not have bounded treewidth seemed interesting to us. We show a natural construction of such patterns in Section \ref{sec:rep}. 

We continue this survey with a result showing that considering some of the structural parameters, that lead to efficient pattern matching algorithms, as general structural parameters of strings, may lead to remarkable results in other apparently unrelated domains. We show in Section \ref{sec:cutwidth} how our results for strings can be used to obtain a state-of-the-art approximation algorithm for computing the {\em cutwidth of graphs}.

We conclude the survey with a series of extensions. We discuss the problem of {\em injective pattern matching} as well as the satisfiability problem for word equations with restricted form.

\section{Basic Definitions}\label{sec:basicDef}

For detailed definitions regarding combinatorics on words we refer to~\cite{Loth97}. \par

We denote our \emph{alphabet} by $\Sigma$, the \emph{empty word} by $\varepsilon$, the set of all non-empty words over $\Sigma$ by $\Sigma^+$,
the set of all words  over $\Sigma$ by $\Sigma^*$, and the \emph{length} of a word $w$ by $|w|$.
$(\Sigma^*,\cdot,\varepsilon)$ is the free monoid over $\Sigma$ with \emph{concatenation} as its binary operation, written~$\cdot$. 
For $w\in \Sigma^*$ and every integers $i,j$ with $1 \leq i\leq  j \leq |w|$, let $w[i..j] = w[i] \cdots w[j]$, where $w[k]$ represents the \emph{letter on position} $k$ and $1\leq k\leq |w|$. A \emph{period} of $w$ is any positive integer $p$ for which $w[i]=w[i+p]$, for all defined positions. Moreover, in this case, $w$ is said to be $p$-periodic. Its \emph{minimal period} is denoted by $per(w)$ and represents the smallest period of $w$. For example, $w = \ta \tb \ta \tc \ta \tb \ta \tc \ta \tb \ta \tc \ta \tb \ta \tc \ta \tb$ has periods $8$ and $4$; in particular, $per(w) = 4$. A word $w$ is called periodic if $per(w)\leq \frac{|w|}{2}$. 

The \emph{concatenation} of $k$ words $w_1, w_2, \ldots,w_k$ is written $\Pi_{i=1,k}w_i$. If $w=w_i$ for all integers $i$ with $1\leq i\leq k$, this  represents the $k$th \emph{power} of $w$, denoted by $w^k$; here, $w$ is a \emph{root} of $w^k$. We can further extend the notion of a power of a word by saying that $w=w[1..per(w)]^{\frac{|w|}{per(w)}}$.
We say that $w$ is \emph{primitive} if it cannot be expressed as a power of exponent $\ell$ of any root, where $\ell$ is an integer with $\ell>1$.
Conversely, if $w=v^\ell$ for some integer $\ell>1$, then $w$ is also called a \emph{repetition}. The infinite repetition $vvv\cdots$ of some word $v$ is denoted $v^\omega$. \par 

For any word $w \in \Sigma^+$ with $w=xyz$, we say that $y$ is a \emph{factor} of $w$. 
If $x$ is empty, then $y$ is also a \emph{prefix} of $w$, while when $z$ is empty, then $y$ is also a \emph{suffix}. 
Whenever we have a factor both as a prefix and as a suffix, the factor is said to be a \emph{border} of the word. 
Furthermore, every word $u=yzx\in \Sigma^+$ is a \emph{conjugate} of $w$. Note that, if $w$ is primitive, so is every conjugate of it. If $w=vu$, then $v^{-1}w=u$. 

Let $X = \{x_1, x_2, x_3, \ldots\}$ and call every $x \in X$ a \emph{variable}. 
For a finite alphabet $\Sigma$ of \emph{terminals} with $\Sigma \cap X = \emptyset$, we define $\pat_{\Sigma} = (X \cup \Sigma)^+$ 
and $\pat = \bigcup_{\Sigma} \pat_{\Sigma}$. Every $\alpha \in \pat$ is a \emph{pattern} and every $w \in \Sigma^*$ is a (\emph{terminal}) \emph{word}. 
Given a word or a pattern $v$, for the smallest sets $B\subseteq\Sigma$ and $Y\subseteq X$ with $v \in (B\cup Y)^*$, we denote $\alphabet(v)=B$ and $\var(v)=Y$. 
For any $x\in \Sigma \cup X$ and $\alpha\in\pat_{\Sigma}$,
$|\alpha|_x$ denotes the number of occurrences of $x$ in $\alpha$; for the sake of convenience, we set $|\alpha|_x = 0$ for every symbol $x$ not in $\Sigma \cup X$. 
For a pattern $\alpha$, we say that $w=\alpha[i..i+|w|]$ is a maximal terminal factor of $\alpha$ if $\alpha[i-1]$ and $\alpha[i+|w|+1]$ are either not defined, or are variables.\par
 
A \emph{substitution} (\emph{for $\alpha$}) is a mapping $h : \var(\alpha) \rightarrow \Sigma^*$. For every $x \in \var(\alpha)$, we say that \emph{$x$ is substituted by $h(x)$} and $h(\alpha)$ denotes the word obtained by substituting every occurrence of a variable $x$ in $\alpha$ by $h(x)$ and leaving the terminals unchanged. We say that the pattern $\alpha$ \emph{matches} $w\in\Sigma^+$ if $h(\alpha)=w$ for some substitution $h : \var(\alpha) \rightarrow \Sigma^*$. Substitutions of the form $h : \var(\alpha) \rightarrow \Sigma^+$, i.\,e., the empty word is excluded from the range of the substitution, are also called \emph{non-erasing}; in order to emphasize that the substitution by the empty word is allowed, we also use the term \emph{erasing} substitution.

\begin{example}
%Let $\beta = x_1  \ta  x_2  \tb x_2  x_1  x_2$ be a pattern and let $u = \tb \ta \tc \tb  \ta  \tb  \tb  \tb  \tb \ta \tc \tb  \tb$ and $v = \ta \tb \ta \ta \tb \tb \ta \tb \ta \tb \ta \tb$ be terminal words. The pattern $\beta$ matches both $u$ and $v$, witnessed by the substitutions $h$ with $h(x_1) = \tb \ta \tc \tb$, $h(x_2) = \tb$ and $g$ with $g(x_1) = g(x_2) = \ta \tb$, respectively. 
Let $\beta = x_1  \ta  x_2  \tb x_2  x_1  x_2$ be a pattern and let $u = \tb \ta \tc \tb  \ta  \tb  \tb  \tb  \tb \ta \tc \tb  \tb$ and $v = \ta \tb \ta \ta \tb \tb \ta \tb \ta \tb \ta \tb$ be terminal words. The pattern $\beta$ matches both $u$ and $v$, witnessed by the substitutions $h$ with $h(x_1) = \tb \ta \tc \tb$, $h(x_2) = \tb$ and $g$ with $g(x_1) = g(x_2) = \ta \tb$, respectively. Moreover, $\beta$ also matches the word $w = \ta  \tc \tb \tb \tc \tb \tc \tb$ by the \emph{erasing} substitution $h$ with $h(x_1) = \varepsilon$, $h(x_2) = \tc \tb$; it can be easily verified that there is no non-erasing substitution that maps $\beta$ to $w$.
\end{example}

The \emph{matching problem}, denoted by $\match$, is to decide for a given pattern $\alpha$ and word $w$, whether there exists a substitution $h$ with $h(\alpha) = w$. The variant where we are only concerned with non-erasing substitutions is called the \emph{non-erasing case} of the matching problem; we also use the term \emph{erasing-case} in order to emphasize that substitution by the empty word is allowed. Another special variant is the \emph{terminal-free case} of the matching problem, where the input patterns are terminal-free, i.\,e., they do not contain any occurrences of terminal symbol. We shall briefly discuss some particularities of these different special cases of the matching problem in Section~\ref{sec:hardness}. Note that in the sections on efficient algorithms, namely Sections \ref{sec:algos}, \ref{sec:local}, and \ref{sec:rep}, we only consider the non-erasing case (with terminal symbols) of the matching problem.
%variants of the matching problem where the sought substitutions which are not allowed to \emph{erase} variables (by mapping them to $\varepsilon$). 
The presented results can easily be generalised to the general setting, but we prefer the respective framework for the ease of the presentation. 

For any $P \subseteq \pat$, the \emph{matching problem for $P$} (or \emph{$\match$ for $P$}, for short) is the matching problem, where the input patterns are from $P$. In the sections of this paper we will introduce and discuss several interesting families of patterns.
\par
As we discuss efficient algorithms, it is important to describe the computational model we use in this work. This is the standard unit-cost RAM with logarithmic word size. Also, all logarithms appearing in our time complexity evaluations are in base 2. For the sake of generality, we assume that whenever we are given as input a word $w\in\Sigma^*$ of length $n$, the symbols of $w$ are in fact integers from $\{1,2,\ldots,n\}$ (i.e., $\Sigma=\alphabet(w)\subseteq \{1,2,\ldots,n\}$), and $w$ is seen as a sequence of integers. This is a common assumption in the area of algorithmics on words (see, e.g., the discussion in~\cite{KaSaBu06}). Clearly, our algorithmic results hold canonically for constant alphabets, as well.\par

\section{The Hardness of the Matching Problem}\label{sec:hardness}

%\todo[inline]{MS: I think that these initial two paragraphs here should be in the paper, since they also cite some people that should be cited in the context of patterns (especially Dominik and Daniel) and make a connection to pattern \emph{languages}. I am not sure, whether this section is the best place for such a paragraph, but I did not see how to incorporate it in a nice way into the introduction (it would somehow break the style of the introduction) or the preliminaries (it is too much giving background to go well with the concise definitions and notations).}

First, we recall that there are several different variants of the matching problem: the most general case (substitution by the empty word and occurrences of terminals in the patterns are possible), the non-erasing case (with terminal symbols), the terminal-free (erasing) case, and finally the terminal free non-erasing case. As we shall see, these differences do not matter too much if we are only concerned with the matching problem of patterns. However, in other contexts of patterns with variables (e.\,g., other decision problems, learning theory), these differences are most crucial and we therefore briefly provide some background.\par
%Patterns can be interpreted as a descriptor of the formal language that contains all words that match the pattern , 
For the class of the so-called \emph{pattern languages}, i.\,e., the sets of all words that match a pattern, the difference between the erasing and the non-erasing case is important, since these classes of formal languages differ quite substantially with respect to basic decision problems. For example, in the non-erasing case, two patterns describe the same language if and only if the patterns are identical (up to a renaming of variables), while it is open whether the equivalence problem is even decidable in the erasing-case (see, e.\,g., Section $6$ in~\cite{MateescuSalomaa1997}, or~\cite{Reidenbach2007}). Moreover, the inclusion problem, which is undecidable for both the erasing and the non-erasing case (see~\cite{JiangEtAl1995,FreydenbergerReidenbach2010b}), can be decided for terminal-free patterns in the erasing case, while for terminal-free non-erasing patterns the decidability status is open (intuitively speaking, this has to do with the fact that avoidability questions of the form ``does pattern $\beta$ necessarily occur in long enough words over a $k$-letter alphabet?'' can be expressed as inclusion for two languages given by terminal-free non-erasing patterns). Finally, also whether patterns (or descriptive patterns) can be inferred from positive data strongly depends on whether the erasing or non-erasing case is considered, or whether or not terminal symbols in the patterns are allowed (see~\cite{Reidenbach2006,rei:dis,FreydenbergerReidenbach2013,FreydenbergerReidenbach2010}).\par
For the matching problem (note that this corresponds to the membership problem for pattern languages), whether we consider erasing or non-erasing substitution, or whether or not we disallow terminal symbols in the patterns, has little impact on its computational hardness. In fact, that the matching problem for patterns with variables is $\npclass$-complete has been independently discovered in different communities and for slightly different problem variants (see, e.\,g., the introductions of~\cite{FernauSchmid2015,FernauEtAl2016} for some remarks on the history of the investigation of the matching problem). \par
%The fact that the matching problem for patterns with variables is $\npclass$-complete has been independently discovered in different communities. 
If we consider the most general case, i.\,e., erasing substitutions and terminals in the patterns, then a hardness-reduction is rather simple. For example, the Boolean formula $$((v_1, v_2, v_3), (v_2, v_4, v_5), (v_3, v_1, v_3), (v_4, v_1, v_2))$$ in $3$-CNF (without negated variables) is $1$-in-$3$ satisfiable (i.\,e., satisfiable with exactly one literal per clause set to \emph{true}) if and only if the following word $w$ is matched by the pattern $\alpha$:%\medskip\\
\begin{center}
\begin{tabular}{llclclclc}
$w$ &$=$ &$\ta$ &$\tb$ &$\ta$ &$\tb$ &$\ta$ &$\tb$ &$\ta$\\
$\alpha$ &$=$ &$\:x_1 x_2 x_3\:$ &$\tb$ &$\:x_2 x_4 x_5\:$ &$\tb$ &$\:x_3 x_1 x_3\:$ &$\tb$ &$\:x_4 x_1 x_2\:$
\end{tabular}
\end{center}
%\medskip\\
%For example, let $$((v_1, v_2, v_3), (v_2, v_4, v_5), (v_3, v_1, v_3), (v_4, v_1, v_2))$$ be a Boolean formula in $3$-CNF (without negated variables). Then this formula is $1$-in-$3$ satisfiable (i.\,e., satisfiable with exactly one literal per clause set to \emph{true}) if and only if the word $\ta \tb \ta \tb \ta \tb \ta$ matches the pattern $x_1 x_2 x_3 \tb x_2 x_4 x_5 \tb x_3 x_1 x_3 \tb x_4 x_1 x_2$. \par
We can further observe that this simple reduction also shows that the matching problem is hard even for binary terminal alphabets and under the restriction that variables are substituted by single symbols (or the empty word) only. This directly raises the questions under which restrictions the matching problem remains hard. For example, a problem instance has a large number of natural parameters (length of the pattern, length of the word, number of variables, number of occurrences per variable, alphabet size, length of words substituted for variables) and in addition to that, it comes in four natural variants resulting from whether we consider the erasing or non-erasing case, and whether or not we allow terminals in the pattern.
 %the problem has four different natural variants, resulting from whether we consider the erasing or non-erasing case, and whether or not we allow terminals in the pattern. 
 In the above reduction, the number of variables, the number of occurrences per each variable and the word length are unbounded. \par
All these numerous restricted problem variants have been thoroughly investigated in~\cite{FernauSchmid2015} and it turns out that the matching problem remains $\npclass$-hard under rather strong restrictions. We cite the following result as an example and refer to~\cite{FernauSchmid2015} for further details.

\begin{theorem}[\cite{FernauSchmid2015}]\label{hardnessExampleTheorem}
The erasing case of the matching problem for patterns with variables is $\npclass$-complete, even if $\Sigma = \{\ta, \tb\}$, every variable has at most $2$ occurrences and every variable can only be substituted by a single symbol or the empty word.
\end{theorem}

This result also holds as stated for terminal-free patterns. In the non-erasing case, however, it holds when the bound on the substitution words is $3$ instead of $1$, and in the non-erasing and terminal-free case the result holds when additionally the bounds on the occurrences per variable and alphabet size are $3$ and $4$, respectively.\par
The only polynomial-time solvable cases of the matching problem obtained by restricting the numerical parameters mentioned above are trivial ones. More precisely, the matching problem can be easily solved for unary alphabets (in this case, we only have to solve an equation in the integers and with integer coefficients, which are given in unary encoding), or if every variable has only one occurrence (the patterns are then \emph{regular}, see Section~\ref{sec:Classes}), or if the number of variables or the length of the input word is bounded by a constant (the former is obvious, while the latter, in the erasing case, requires a slightly more careful argument~\cite{gei:lea}). \par
In particular, this also points out that Theorem~\ref{hardnessExampleTheorem} describes some kind of dichotomy, i.\,e., if we would further restrict the alphabet size, or the maximum number of occurrences per variable to $1$, then we would obtain a polynomial-time solvable variant (even if all other parameters are unrestricted); similarly, if we allow variables to be substituted by single symbols only, but not the empty word, then the matching problem becomes efficiently solvable as well (regardless of the alphabet size).\par
Generally, by brute-force algorithms, the matching problem can be solved in time $|\alpha|^{\bigO(|w|)}$ or $|w|^{\bigO(|\alpha|)}$, making it polynomial-time solvable provided that there is a constant upper bound on $|w|$ or $|\alpha|$ (in fact, a bound on $|\var(\alpha)|$ is sufficient). However, this constant upper bound occurs in the exponent, which means that even for rather low such bounds, say $7$, the corresponding polynomial-time algorithms are most likely impractical for larger problem instances. This leads to the question whether exponential-time algorithms are possible whose running-times are such that the exponential part \emph{exclusively} depends on, say $|\var(\alpha)|$, but not on $|w|$, i.\,e., running-times of the form $f(|\var(\alpha)|) \times g(|\alpha|, |w|)$, where $g$ is a polynomial and $f$ is some \emph{computable} function (exponential, or even double-exponential etc.). Such a running-time is polynomial for upper bounded $|\var(\alpha)|$, but the degree of the polynomial is always the same independent from the actual upper bound. If a problem has an algorithm with such a running-time, then it is called \emph{fixed-parameter tractable} (with respect to the bounded parameter); see the textbooks~\cite{DowneyFellows2013,FlumGrohe06} for more information on parameterised complexity. Whether the matching problem for patterns with variables allows fixed-parameter tractability for some parameters has been thoroughly investigated in~\cite{FernauEtAl2016}. Although there are some more or less trivial cases of fixed-parameter tractability, the main insight provided by~\cite{FernauEtAl2016} is of a negative nature and can be summarised in the following way.

\begin{theorem}[\cite{FernauEtAl2016}]\label{FPThardnessExampleTheorem}
All variants of the matching problem parameterised by $|\alpha|$ are $\wclass[1]$-hard. The erasing case of the matching problem parameterised by $|w|$ is $\wclass[1]$-hard.\footnote{Problems that are hard for the parameterised complexity class $\wclass[1]$ are strongly believed to be not fixed-parameter tractable.}
\end{theorem}

Note that since $|\alpha|$ and $|w|$ are rather general parameters, this result covers other parameters as well, e.\,g., $|\Sigma|$ or $|\var(\alpha)|$. In the non-erasing case, $|w|$ is an upper bound for $|\var(\alpha)|$; thus, treating $|w|$ as a parameter means that $|\var(\alpha)|$ is also a parameter and therefore the matching problem is fixed-parameter tractable by the obvious brute-force algorithm. We refer to~\cite{FernauEtAl2016} for further such simple fixed-parameter tractable case. \par
Consequently, even strong restrictions of the obvious numerical parameters of instances of the matching problem, i.\,e., number of variables, alphabet size, occurrences per variable etc., does not yield interesting efficiently matchable subclasses of patterns with variables. However, as discussed in the next section, looking deeper into the structure of patterns will help.

\section{Structural Restrictions for Patterns}\label{sec:Classes}

%THINGS TO ADD
%Meta-theorem with a sketch of the graph encoding.
%
%An overview of pattern classes with bounded treewidth
%
%Proof that local patterns have bounded treewidth? (is this published somewhere?)

From an intuitive point of view it is clear that not only the mere length of a pattern or the number of its variables should have an impact on the matching complexity, but also the actual order of the variables. For example, it has been observed rather early in~\cite{shi:pol2} that if the variable occurrences in the patterns are sorted, e.\,g., as in $x_1 \ta x_1 x_2 x_2 \ta \tb x_2 x_2 \ta x_3 x_4 \tc x_4$, then they can be matched efficiently ``from-left-to-right'' (more precisely, it is observed in~\cite{shi:pol2} that matching such patterns can be done in logarithmic space).\par
A systematic investigation of such structural restrictions has been done in the last decade and numerous efficiently matchable subclasses of patterns have been found. In the following, we first present a unifying approach based on graph morphisms and the concept of treewidth. Then, we define and summarise several structural parameters for patterns and respective subclasses of patterns.

\subsection{Pattern Matching by Graph Morphisms}\label{sec:treewidth}

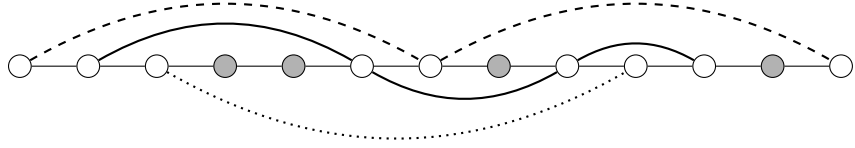
\begin{figure}
%\begin{center}
\centering
\begin{tikzpicture}[nodes={draw, circle, inner sep=3pt}, scale=0.5]
\coordinate (xshift) at (1.8,0);
\coordinate (yshift) at (0,2);

\node (1) at (0,0) {};
\node (2) at ($(1) + (xshift)$) {};
\node (3) at ($(2) + (xshift)$) {};
\node[fill=black!30] (3a) at ($(3) + (xshift)$) {};
\node[fill=black!30] (3b) at ($(3a) + (xshift)$) {};
\node (4) at ($(3b) + (xshift)$) {};
\node (5) at ($(4) + (xshift)$) {};
\node[fill=black!30] (5a) at ($(5) + (xshift)$) {};
\node (6) at ($(5a) + (xshift)$) {};
\node (7) at ($(6) + (xshift)$) {};
\node (8) at ($(7) + (xshift)$) {};
\node[fill=black!30] (8a) at ($(8) + (xshift)$) {};
\node (9) at ($(8a) + (xshift)$) {};

\draw (1) edge[] (2);
\draw (2) edge[] (3);
\draw (3) edge[] (3a);
\draw (3a) edge[] (3b);
\draw (3b) edge[] (4);
\draw (4) edge[] (5);
\draw (5) edge[] (5a);
\draw (5a) edge[] (6);
\draw (6) edge[] (7);
\draw (7) edge[] (8);
\draw (8) edge[] (8a);
\draw (8a) edge[] (9);

\draw (1) edge[thick, bend left, dashed] (5);
%\draw (1) edge[thick, bend right=20, dashed] (9);
\draw (5) edge[thick, bend left, dashed] (9);

\draw (2) edge[thick, bend left] (4);
%\draw (2) edge[thick, bend left=18] (6);
%\draw (2) edge[thick, bend left=25] (8);
\draw (4) edge[thick, bend right] (6);+
%\draw (4) edge[thick, bend right=25] (8);
\draw (6) edge[thick, bend left] (8);

\draw (3) edge[thick, bend right, dotted] (7);
\end{tikzpicture}
%\end{center}
\caption{The standard graph representation $G^{\pattern}_{\alpha}$ for $\alpha = x_1 x_2 x_3 \tb \tb x_2 x_1 \ta x_2 x_3 x_2 \tc x_1$; the dashed, straight and dotted equality edges correspond to occurrences of $x_1$, $x_2$ and $x_3$, respectively; the grey vertices correspond to occurrences of terminal symbols.}
\label{fig:standardGraphRepresentations}
\end{figure}

The following general framework for matching patterns with variables has been developed in~\cite{rei:patIaC}. For a pattern $\alpha \in (X \cup \Sigma^*)$, the \emph{standard graph representation} of $\alpha$ is the undirected graph $\mathcal{G}^{\pattern}_{\alpha} = (V_{\alpha}, E_{\alpha})$, where $V_{\alpha} = \{1, 2, \ldots, |\alpha|\}$ and $E_{\alpha} = E^{\equality}_{\alpha} \cup E^{\neighbour}_{\alpha}$ with $E^{\equality}_{\alpha} = \{\{i, i+1\} \mid 1 \leq i \leq |\alpha-1|\}$ being the set of \emph{neighbour edges} and $E^{\equality}_{\alpha} = \{\{i, j\} \mid \alpha[i..j] = x \beta x, x \in X, |\beta|_x = 0\}$ being the set of \emph{equality edges} (see Figure~\ref{fig:standardGraphRepresentations} for an illustration). 

In a similar way, we can also encode words $w \in \Sigma^*$ as graph structures $\mathcal{G}^{\word}_{w}$, where every factor $w[i..j]$ of $w$ is represented by a vertex $(i, j)$, equality edges are drawn between $(i, j)$ and $(i', j')$ if $w[i..j] = w[i'..j']$, and neighbour edges if $j + 1 = i'$. It has been shown in~\cite{rei:patIaC} that $\alpha$ matches $w$ if and only if there is a graph morphism from $\mathcal{G}^{\pattern}_{\alpha}$ to $\mathcal{G}^{\word}_{w}$. Moreover, the concept of the treewidth for graphs now also applies to patterns (i.\,e., the treewidth of a pattern is the treewidth of its standard graph representation), which is of relevance since the graph morphism problem can be solved in polynomial-time provided that the source graphs have bounded treewidth.\footnote{See~\cite{DowneyFellows2013,FlumGrohe06} for a formal definition of the treewidth.} Consequently, we can conclude the following algorithmic meta-theorem. 

\begin{theorem}[\cite{rei:patIaC}]\label{metaTheorem}
If a class $P$ of patterns has bounded treewidth, then the matching problem for $P$ can be solved in polynomial-time.
\end{theorem}

Due to the generality of the statement of Theorem~\ref{metaTheorem}, the polynomial-time matching algorithm that it implies is of little practical value, even for rather simple classes of patterns. On the other hand, its theoretical relevance is demonstrated by the fact that it covers almost all known classes of patterns with a polynomial-time matching problem.\footnote{See Section~\ref{sec:rep} for the respective exceptions.} After an additional remark regarding~\cite{rei:patIaC}, we shall briefly define and compare those efficiently matchable classes of patterns in the next subsection.

%We conclude this section by an additional remark, and in the next subsection, we shall briefly define and compare those efficiently matchable classes of patterns.

\begin{remark}
%In order to briefly sketch the general idea of~\cite{rei:patIaC} above, we omit some details that are negligible here. 
Technically, the matching problem reduces to the morphism problem for (simple) relational structures instead of undirected graphs. However, since we are here only interested in the treewidth of these structures, we can as well only talk about the underlying undirected graphs. \par
Moreover, the actual meta-theorem of~\cite{rei:patIaC} is stronger in the sense that there the treewidth of patterns is not defined with respect to the standard graph representation, but with respect to a slightly more general graph representations (i.\,e., we allow any way of drawing the equality edges as long as all vertices corresponding to the same variable form a connected component). 
\end{remark}

\subsection{Efficiently Matchable Classes of Patterns}\label{sec:graphs}

The most obvious way to restrict patterns is to limit their number of (repeated) variables or the number of occurrences per variable. In this regard, let $\var_k$ and $\repeatedvar_k$ be the class of patterns with at most $k$ variables and with at most $k$ repeated variables, respectively. Due to Theorem~\ref{hardnessExampleTheorem}, we already know that bounding the number of occurrences per variable does not in general yield polynomial-time matchable classes. The only exception are patterns with at most one occurrence per variable, which are called \emph{regular} patterns and are denoted by $\regular$, e.\,g., $x_1 \ta x_2 \tb \ta \tc x_3 \ta$ is a regular pattern. Regular patterns have been first considered in~\cite{shi:pol2} and their name is motivated by the fact that the corresponding pattern languages are regular languages. \par
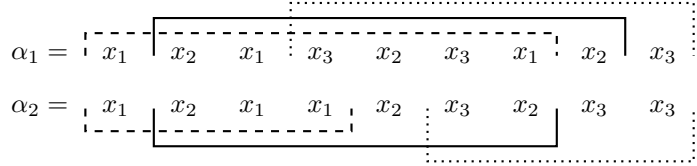
\begin{figure}
\begin{center}
\begin{tikzpicture}

\coordinate (coord0) at (0,0);
\coordinate (coord1) at ($(coord0) + (0.9,0)$);
\coordinate (coord2) at ($(coord1) + (0.9,0)$);
\coordinate (coord3) at ($(coord2) + (0.9,0)$);
\coordinate (coord4) at ($(coord3) + (0.9,0)$);
\coordinate (coord5) at ($(coord4) + (0.9,0)$);
\coordinate (coord6) at ($(coord5) + (0.9,0)$);
\coordinate (coord7) at ($(coord6) + (0.9,0)$);
\coordinate (coord8) at ($(coord7) + (0.9,0)$);

\draw ($(coord0) + (-2,-4.5)$) node {$\alpha_1 =$};
\draw ($(coord0) + (-1,-4.5)$) node {$x_1$};
\draw ($(coord1) + (-1,-4.5)$) node {$x_2$};
\draw ($(coord2) + (-1,-4.5)$) node {$x_1$};
\draw ($(coord3) + (-1,-4.5)$) node {$x_3$};
\draw ($(coord4) + (-1,-4.5)$) node {$x_2$};
\draw ($(coord5) + (-1,-4.5)$) node {$x_3$};
\draw ($(coord6) + (-1,-4.5)$) node {$x_1$};
\draw ($(coord7) + (-1,-4.5)$) node {$x_2$};
\draw ($(coord8) + (-1,-4.5)$) node {$x_3$};

%\draw[white] ($(coord0) + (-1,-4.5) + (-0.2, 0)$) -- ($(coord0) + (-1,-4.5) + (-0.2, 0.3)$) -- ($(coord6) + (-1,-4.5) + (0.2, 0.3)$) -- ($(coord6) + (-1,-4.5) + (0.2, 0)$);
%\draw[white] ($(coord1) + (-1,-4.5) + (-0.2, 0)$) -- ($(coord1) + (-1,-4.5) + (-0.2, 0.5)$) -- ($(coord7) + (-1,-4.5) + (0.2, 0.5)$) -- ($(coord7) + (-1,-4.5) + (0.2, 0)$);
%\draw[white] ($(coord3) + (-1,-4.5) + (-0.2, 0)$) -- ($(coord3) + (-1,-4.5) + (-0.2, 0.7)$) -- ($(coord8) + (-1,-4.5) + (0.2, 0.7)$) -- ($(coord8) + (-1,-4.5) + (0.2, 0)$);
\draw[dashed, thick] ($(coord0) + (-1.2,-4.5) + (-0.2, 0)$) -- ($(coord0) + (-1.2,-4.5) + (-0.2, 0.3)$) -- ($(coord6) + (-1,-4.5) + (0.4, 0.3)$) -- ($(coord6) + (-1,-4.5) + (0.4, 0)$);
\draw[thick] ($(coord1) + (-1,-4.5) + (-0.4, 0)$) -- ($(coord1) + (-1,-4.5) + (-0.4, 0.5)$) -- ($(coord7) + (-1,-4.5) + (0.4, 0.5)$) -- ($(coord7) + (-1,-4.5) + (0.4, 0)$);
\draw[dotted, thick] ($(coord3) + (-1,-4.5) + (-0.4, 0)$) -- ($(coord3) + (-1,-4.5) + (-0.4, 0.7)$) -- ($(coord8) + (-1,-4.5) + (0.4, 0.7)$) -- ($(coord8) + (-1,-4.5) + (0.4, 0)$);

%\draw[black] ($(coord8) + (0.5,-4.5)$) node {$\scd(\alpha_1) = 3$};

\draw ($(coord0) + (-2,-5.2)$) node {$\alpha_2 =$};
\draw ($(coord0) + (-1,-5.2)$) node {$x_1$};
\draw ($(coord1) + (-1,-5.2)$) node {$x_2$};
\draw ($(coord2) + (-1,-5.2)$) node {$x_1$};
\draw ($(coord3) + (-1,-5.2)$) node {$x_1$};
\draw ($(coord4) + (-1,-5.2)$) node {$x_2$};
\draw ($(coord5) + (-1,-5.2)$) node {$x_3$};
\draw ($(coord6) + (-1,-5.2)$) node {$x_2$};
\draw ($(coord7) + (-1,-5.2)$) node {$x_3$};
\draw ($(coord8) + (-1,-5.2)$) node {$x_3$};

%\draw[white] ($(coord0) + (-1,-5.2) + (-0.2, 0)$) -- ($(coord0) + (-1,-5.2) + (-0.2, -0.3)$) -- ($(coord3) + (-1,-5.2) + (0.2, -0.3)$) -- ($(coord3) + (-1,-5.2) + (0.2, 0)$);
%\draw[white] ($(coord1) + (-1,-5.2) + (-0.2, 0)$) -- ($(coord1) + (-1,-5.2) + (-0.2, -0.5)$) -- ($(coord6) + (-1,-5.2) + (0.2, -0.5)$) -- ($(coord6) + (-1,-5.2) + (0.2, 0)$);
%\draw[white] ($(coord5) + (-1,-5.2) + (-0.2, 0)$) -- ($(coord5) + (-1,-5.2) + (-0.2, -0.7)$) -- ($(coord8) + (-1,-5.2) + (0.2, -0.7)$) -- ($(coord8) + (-1,-5.2) + (0.2, 0)$);
\draw[dashed, thick] ($(coord0) + (-1,-5.2) + (-0.4, 0)$) -- ($(coord0) + (-1,-5.2) + (-0.4, -0.3)$) -- ($(coord3) + (-1,-5.2) + (0.4, -0.3)$) -- ($(coord3) + (-1,-5.2) + (0.4, 0)$);
\draw[thick] ($(coord1) + (-1,-5.2) + (-0.4, 0)$) -- ($(coord1) + (-1,-5.2) + (-0.4, -0.5)$) -- ($(coord6) + (-1,-5.2) + (0.4, -0.5)$) -- ($(coord6) + (-1,-5.2) + (0.4, 0)$);
\draw[dotted, thick] ($(coord5) + (-1,-5.2) + (-0.4, 0)$) -- ($(coord5) + (-1,-5.2) + (-0.4, -0.7)$) -- ($(coord8) + (-1,-5.2) + (0.4, -0.7)$) -- ($(coord8) + (-1,-5.2) + (0.4, 0)$);

%\draw ($(coord8) + (0.5,-5.2)$) node {$\scd(\alpha_2) = 2$};

\end{tikzpicture}
\end{center}
\caption{Two pattern $\alpha_1$ and $\alpha_2$ with $\scd(\alpha_1) = 3$ and $\scd(\alpha_2) = 2$. The scopes of variable $x_1$ (dashed line), $x_2$ (straight line) and $x_3$ (dotted line) are highlighted.
%; the dashed, straight and dotted equality edges correspond to occurrences of $x_1$, $x_2$ and $x_3$, respectively; the grey vertices correspond to occurrences of terminal symbols.
}
\label{fig:scdIllustration}
\end{figure}

Next, we define the so-called scope coincidence degree (see~\cite{rei:patIaC}). For every $y \in \var(\alpha)$, the \emph{scope of $y$ in $\alpha$} is defined by $\scope_{\alpha}(y) = \{i, i+1,\ldots, j\}$, where $i$ is the leftmost and $j$ the rightmost occurrence of $y$ in $\alpha$. The scopes of some variables $y_1, y_2,\ldots, y_k \in \var(\alpha)$ \emph{coincide in $\alpha$} if $\bigcap_{1 \leq i \leq k} \scope_{\alpha}(y_i) \neq \emptyset$. By $\scd(\alpha)$, we denote the \emph{scope coincidence degree} of $\alpha$, which is the maximum number of variables in $\alpha$ such that their scopes coincide, and by $\scd_k$, we denote the class of patterns with scope coincidence degree of at most $k$. See Figure~\ref{fig:scdIllustration} for an example of the scope coincidence degree. An important special class is $\scd_1$, which has been first introduced in~\cite{shi:pol2} as the class of \emph{non-cross} patterns (denoted by $\noncross$). Intuitively speaking, the variables in non-cross patterns are sorted, e.\,g., $x_1 \ta x_1 x_2 \tb \ta x_2 \tc x_3 \ta x_3 x_3$. 

%An important special class is $\scd_1$, which is also called the class of \emph{non-cross} patterns (denoted by $\noncross$). Intuitively speaking, the variables in non-cross patterns are sorted, e.\,g., $x_1 \ta x_1 x_2 \tb \ta x_2 \tc x_3 \ta x_3 x_3$. 

Next, we define the locality number, which is a general string-parameter, and which has been first introduced in~\cite{DayEtAl2017}. A word is $k$-local if there exists an order of its symbols such that, if we {\em mark} the symbols in the respective order (which is called a \emph{marking sequence}), at each stage there are at most $k$ contiguous blocks of marked symbols in the word. This $k$ is called the {\em marking number} of that marking sequence. The \emph{locality number} of a word is the smallest $k$ for which that word is $k$-local, or, in other words, the minimum marking number over all marking sequences. For example, the marking sequence $\sigma = (\ta, \tg, \tc)$ marks $w = \ta \tg \ta \tg \tc \ta \tc$ as follows (marked blocks are illustrated by overlines): $\ta \tg \ta \tg \tc \ta \tc$, $\overline{\ta} \tg \overline{\ta} \tg \tc \overline{\ta} \tc$, $\overline{\ta \tg \ta \tg} \tc \overline{\ta} \tc$, $\overline{\ta \tg \ta \tg \tc \ta \tc}$; thus, the marking number of $\sigma$ is $3$. In fact, all marking sequences for $w$ have a marking number of $3$, except $(\tg, \ta, \tc)$, for which it is $2$: $\ta \overline{\tg} \ta \overline{\tg} \tc \ta \tc$, $\overline{\ta \tg \ta \tg} \tc \overline{\ta} \tc$, $\overline{\ta \tg \ta \tg \tc \ta \tc}$. Thus, the locality number of $w$, denoted by $\loc(w)$, is $2$. When we measure the locality number for patterns, we simply ignore all terminal symbols, e.\,g., $\loc(\ta \tb x_1 x_2 \ta x_1 x_2 \tc x_3 x_1 \ta x_3) = \loc(x_1 x_2 x_1 x_2 x_3 x_1 x_3) = 2$. The class of patterns with locality number at most $k$ is denoted by $\loc_k$.

The next classes have been first considered in~\cite{rei:patIaC} and are based on possible nesting structures of variables. For a pattern $\alpha$, we call two variables $x, y \in \var(\alpha)$ \emph{entwined} if $\alpha$ contains $xyxy$ or $yxyx$ as a subsequence. A pattern $\alpha$ is \emph{nested}, if no two variables in $\alpha$ are entwined; the class of nested patterns is denoted by $\nested$. A proper subclass of $\nested$, considered in~\cite{DayEtAl2017}, are the so-called \emph{strongly} nested patterns (denoted by $\stronglynested$), which are inductively defined as follows: any pattern $\alpha \in \var_1$ is strongly nested; if $\alpha_1$ and $\alpha_2$ are strongly nested and variable-disjoint patterns, $x$ is a variable not in $\var(\alpha_1) \cup \var(\alpha_2)$ and $\beta_1, \beta_2 \in (\{x\} \cup \Sigma)^*$, then $\alpha_1 \alpha_2$ and $\beta_1 \alpha_1 \beta_2$ are strongly nested patterns. For example, the pattern $\alpha = x_1 x_2 \ta x_2 x_1 \tb x_3 x_4 \ta x_3$ is strongly nested, whereas $\alpha x_1$ is nested, but not strongly nested anymore.\par
\begin{figure}
%\begin{center}
\centering
\begin{tikzpicture}[nodes={draw, circle, inner sep=2.5pt}, scale=0.5]
\coordinate (xshift) at (1.4,0);
\coordinate (yshift) at (0,2);

\node (1) at (0,0) {};
\node (2) at ($(1) + (xshift)$) {};
\node (3) at ($(2) + (xshift)$) {};
%\node (4) at ($(3) + (xshift)$) {};
%\node (5) at ($(3) + (xshift)$) {};
\node (6) at ($(3) + (xshift)$) {};
\node (7) at ($(6) + (xshift)$) {};
\node (8) at ($(7) + (xshift)$) {};
\node (9) at ($(8) + (xshift)$) {};
\node (10) at ($(9) + (xshift)$) {};
\node (10a) at ($(10) + (xshift)$) {};
\node (11) at ($(10a) + (xshift)$) {};
\node (12) at ($(11) + (xshift)$) {};
\node (13) at ($(12) + (xshift)$) {};
\node (14) at ($(13) + (xshift)$) {};
\node (14a) at ($(14) + (xshift)$) {};
\node (15) at ($(14a) + (xshift)$) {};
\node (16) at ($(15) + (xshift)$) {};

\draw (1) edge[] (2);
\draw (2) edge[] (3);
\draw (3) edge[] (6);
%\draw (4) edge[] (5);
%\draw (5) edge[] (6);
\draw (6) edge[] (7);
\draw (7) edge[] (8);
\draw (8) edge[] (9);
\draw (9) edge[] (10);
\draw (10) edge[] (10a);
\draw (10a) edge[] (11);
\draw (11) edge[] (12);
\draw (12) edge[] (13);
\draw (13) edge[] (14);
\draw (14) edge[] (14a);
\draw (14a) edge[] (15);
\draw (15) edge[] (16);

\draw (1) edge[thick, bend left] (7);
\draw (2) edge[thick, bend left] (6);
%\draw (3) edge[thick, bend left] (4);
%\draw (5) edge[thick, bend left] (6);
\draw (6) edge[thick, bend right] (9);
\draw (8) edge[thick, bend left] (12);
\draw (10) edge[thick, bend left] (11);
\draw (11) edge[thick, bend right] (13);
\draw (12) edge[thick, bend left] (16);
\draw (14) edge[thick, bend left] (15);

\end{tikzpicture}
%\end{center}
\caption{The standard graph representation $G^{\pattern}_{\alpha}$ for $\alpha = x_1 x_3 x_4 x_3 x_1 x_2 x_3 x_5 \tb x_5 x_2 x_5 x_6 \ta x_6 x_2$. By definition, $\alpha$ is mildly entwined. Furthermore, since no vertex is completely ``surrounded'' by edges, the shown embedding is outer-planar.
%; the dashed, straight and dotted equality edges correspond to occurrences of $x_1$, $x_2$ and $x_3$, respectively; the grey vertices correspond to occurrences of terminal symbols.
}
\label{fig:mildlyEntwinedPattern}
\end{figure}
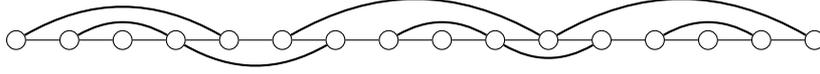

If, for every $x, y \in \var(\alpha)$, $\alpha = \beta x \gamma_1 y \gamma_2 x \gamma_3 y \delta$ implies $\gamma_2 = \varepsilon$, then $\alpha$ is called \emph{closely entwined}, and a pattern $\alpha$ is \emph{mildly entwined} if it is closely entwined and, for every $x \in \var(\alpha)$, if $\alpha = \beta x \gamma x \delta$ with $|\gamma|_x = 0$, then $\gamma$ is nested. We denote the class of mildly entwined patterns by $\mildlyentwined$. The main motivation for the somewhat peculiar class of mildly entwined patterns is that mildly entwined patterns are exactly those patterns that have a standard graph representation that is outer-planar (see~\cite{rei:patIaC}).\footnote{A graph is outer-planar if it has a planar embedding with all vertices lying on the outer face.} It is known that outer-planar graphs have a rather low treewidth of at most $2$. Since the concept of outer-planarity generalises to $k$-outer-planarity and $k$-outer-planar graphs have a treewidth of at most $3k-1$, we can also define the classes $\outp_k$ of $k$-outer-planar patterns (i.\,e., their standard graph representation is $k$-outer-planar). In this regard note that $\outp_1 = \mildlyentwined$. See Figure~\ref{fig:mildlyEntwinedPattern} for an example of a mildly-entwined pattern.

\begin{figure}
\begin{center}
\begin{tikzpicture}
\tikzstyle{every node}=[draw, outer sep=0.1cm]

\coordinate (hshift) at (1.4, 0);
\coordinate (vshift) at (0, 1);

\node (nc) at (0,0) {$\noncross$};
\node (reg) at ($(nc)-(hshift)$) {$\regular$};
\node (var1) at ($(nc)-(vshift)+0*(hshift)$) {$\var_1$};
\node (var2) at ($(var1)+2*(hshift)$) {$\var_2$};
\node (var3) at ($(var2)+2*(hshift)$) {$\var_3$};
\node (var4) at ($(var3)+(hshift)$) {$\var_4$};
\node[draw=none] (var5dots) at ($(var4)+(hshift)$) {$\ldots$};
\node (repv1) at ($(nc)-2*(vshift)+0*(hshift)$) {$\repeatedvar_1$};
\node (repv2) at ($(repv1)+2*(hshift)$) {$\repeatedvar_2$};
\node (repv3) at ($(repv2)+2*(hshift)$) {$\repeatedvar_3$};
\node (repv4) at ($(repv3)+(hshift)$) {$\repeatedvar_4$};
\node[draw=none] (repv5dots) at ($(repv4)+(hshift)$) {$\ldots$};
\node (scd2) at ($(nc)+2*(hshift)$) {$\scd_2$};
\node (scd3) at ($(scd2)+2*(hshift)$) {$\scd_3$};
\node (scd4) at ($(scd3)+(hshift)$) {$\scd_4$};
\node[draw=none] (scd5dots) at ($(scd4)+(hshift)$) {$\ldots$};
\node (loc1) at ($(nc)+3*(vshift)-0*(hshift)$) {$\loc_1$};
\node (loc2) at ($(loc1)+(hshift)$) {$\loc_2$};
\node (loc3) at ($(loc2)+(hshift)$) {$\loc_3$};
\node (loc4) at ($(loc3)+(hshift)$) {$\loc_4$};
\node[draw=none] (loc5dots) at ($(loc4)+(hshift)$) {$\ldots$};
\node (snest) at ($(loc1)+(hshift)-(vshift)$) {$\stronglynested$};
\node (nest) at ($(snest)+(hshift)$) {$\nested$};
\node (ment) at ($(nest)+(hshift)$) {$\mildlyentwined$};
\node (outp2) at ($(ment)+(hshift)$) {$\outp_2$};
\node (outp3) at ($(outp2)+(hshift)$) {$\outp_3$};
\node[draw=none] (outp4dots) at ($(outp3)+(hshift)$) {$\ldots$};
\node (pat) at ($(scd5dots)+(vshift)+1*(hshift)$) {$\allpatterns$};

\draw[->] (nc) edge (scd2);
\draw[->] (nc) edge (loc1);
\draw[->] (reg) edge (nc);
\draw[->] (var1) edge (nc);
\draw[->, bend right = 30, in=-135] (reg) edge (repv1);

\draw[->] (var1) edge (var2);
\draw[->] (var2) edge (var3);
\draw[->] (var3) edge (var4);
\draw[->] (var4) edge (var5dots);
\draw[->, out=0, in=230] (var5dots) edge (pat);
\draw[->] (var1) edge (repv1);
\draw[->] (var2) edge (repv2);
\draw[->] (var3) edge (repv3);
\draw[->] (var4) edge (repv4);
\draw[->] (var2) edge (scd2);
\draw[->] (var3) edge (scd3);
\draw[->] (var4) edge (scd4);

\draw[->] (repv1) edge (repv2);
\draw[->] (repv2) edge (repv3);
\draw[->] (repv3) edge (repv4);
\draw[->] (repv4) edge (repv5dots);
\draw[->, out=0, in=-90] (repv5dots) edge (pat);

\draw[->] (scd2) edge (scd3);
\draw[->] (scd3) edge (scd4);
\draw[->] (scd4) edge (scd5dots);
\draw[->, out=0, in=200] (scd5dots) edge (pat);

\draw[->] (loc1) edge (loc2);
\draw[->] (loc2) edge (loc3);
\draw[->] (loc3) edge (loc4);
\draw[->] (loc4) edge (loc5dots);
\draw[->, bend left = 25, in=120] (loc5dots) edge (pat);

\draw[->, bend right = 30, in=-145] (loc1) edge (snest);
\draw[->] (snest) edge (nest);
\draw[->] (nest) edge (ment);

\draw[->] (ment) edge (outp2);
\draw[->] (outp2) edge (outp3);
\draw[->] (outp3) edge (outp4dots);
\draw[->, out=0] (outp4dots) edge (pat);

\draw[->, bend right = 5, out=-30] (repv1) edge (nest);

\draw[->, bend right = 15, out=-30, in=-180] (repv2) edge (outp2);

\draw[->, bend left = 15, out=-20] (scd2) edge (outp2);

\end{tikzpicture}
\end{center}
\caption{An overview of efficiently matchable classes of patterns. By $A \rightarrow B$, we denote $A \subset B$; pairs without arrow are incomparable. Note that $\noncross = \scd_1$ and $\mildlyentwined = \outp_1$.}
\label{mainDiagramFigure}
\end{figure}
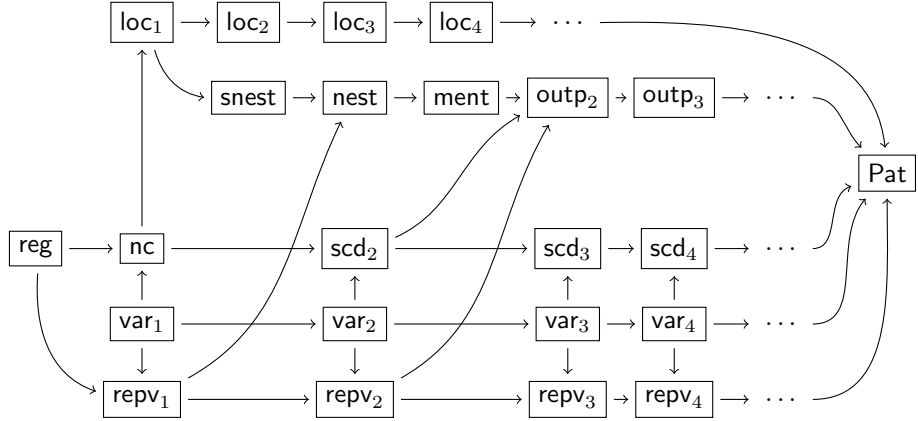

It can be easily verified that all of the pattern classes defined above have bounded treewidth; thus, by application of Theorem~\ref{metaTheorem}, they can be matched efficiently. For some of them this upper bound on the treewidth is rather low (e.\,g., $\regular$, $\noncross$, $\mildlyentwined$), while for those classes obtained by bounding a structural parameter, e.\,g., $\repeatedvar_k$, $\scd_k$, $\loc_k$, the bound on the treewidth also grows with this parameter. Figure~\ref{mainDiagramFigure} shows how these pattern classes relate to each other and how they form infinite hierarchies within the class of all patterns (denoted by $\allpatterns$).

%\begin{tabular}{ll}
%$\regular$ & Regular patterns in which every variable has just one occurrence,\\ & e.\,g., $x_1 \ta x_2 \tb \ta \tc x_3 \ta$,\\
%$\noncross$ & Non-cross patterns in which all variables are sorted, \\ & e.\,g., $x_1 \ta x_1 x_2 \tb \ta x_2 \tc x_3 \ta x_3 x_3$,\\
%$\stronglynested$ & strongly nested patterns,\\
%$\nested$ & nested patterns,\\
%$\mildlyentwined$ & mildly entwined patterns,\\
%$\var_k$ & patterns with at most $k$ variables,\\
%$\repeatedvar_k$ & patterns with at most $k$ repeated variables,\\
%$\scd_k$ & patterns with scope coincidence degree of at most $k$,\\
%$\loc_k$ & $k$-local patterns,\\
%$\outp_k$ & patterns with $k$-outerplanar standard graph representation,\\
%$\treewidth_k$ & patterns with standard graph representation of treewidth $k$,\\
%$\Pi_{*}$ & $\bigcup^{\infty}_{i = 1} \Pi_{k}$, where $\Pi \in \{\var, \repeatedvar, \scd, \loc, \outp, \treewidth\}$,\\
%$\allpatterns$ & all patterns.
%\end{tabular}

%\subsection{Results}
%
%\begin{proposition}
%$\mildlyentwined \subseteq \treewidth_2$ and, for every $k \geq 1$ and for every $\Pi \in \{\var, \repeatedvar, \scd, \loc, \outp\}$, $\Pi_k \subseteq \treewidth_{\bigO(k)}$.
%\end{proposition}
%
%\begin{proposition}
%The subset relations of Figure~\ref{mainDiagramFigure} hold.
%\end{proposition}

In a sense, Figure~\ref{mainDiagramFigure} is a ``tractability map'' for the matching problem of patterns with variables. For the classes that have low treewidth, we can expect matching algorithms that are rather efficient. On the other hand, these classes are quite restricted (compared to the full class of patterns) and are most likely only applicable for very special pattern matching tasks. An obvious approach to matching general patterns would be to first perform a preprocessing that identifies a ``low class'' of the tractability map that contains the input pattern and then uses the most efficient algorithm for matching it. In this regard, it is even an asset that most of the different efficiently matchable classes and hierarchies of classes are incomparable: it is possible that an input pattern has a very large locality number of $100$, but can nevertheless be matched efficiently, because its standard graph representation is $2$-outerplanar; on the other hand, a pattern could have a large scope coincidence degree and a large number of variables, but at the same time a very low locality number. It might even be a worthwhile research task to experimentally analyse a large corpus of (random) patterns with respect to the classes of the tractability map in which they are contained.

\begin{remark}\label{no-fpt}
Bounding the structural parameters defined above yields polynomial-time matchable classes of patterns; thus, the question arises whether the matching problem is also fixed-parameter tractable with respect to those parameters. However, Theorem~\ref{FPThardnessExampleTheorem} already states that this is most likely not the case for parameter $|\var(\alpha)|$, and since $|\var(\alpha)|$ is an upper bound for the number of repeated variables, the scope coincidence degree, the outer-planarity and the locality number of $\alpha$, it is also highly unlikely that we can achieve fixed-parameter tractability with respect to those parameters.
\end{remark}

\subsection{Computing Structural Parameters for Patterns}\label{struct}

Since the structural restrictions of patterns surveyed above are all meant to be exploited algorithmically, the task of checking them (or computing the respective parameters) is an important issue. In this regard, note that in general computing the treewidth of a graph is an $\npclass$-hard problem and it is also not known whether it can be computed efficiently for standard graph representations of patterns. This also emphasises the importance of \emph{easily computable} parameters that are bounding the treewidth of a pattern and also points out why the value of Theorem~\ref{metaTheorem} is of a theoretical nature that provides guidance in finding such restrictions with higher practical relevance. Restrictions like the regularity, the non-cross condition, number of (repeated) variables and the different nesting properties can be easily checked for. Moreover, also the scopes of a pattern and therefore its scope coincidence degree can be efficiently computed, and the smallest $k$ for which a graph is $k$-outerplanar can also be computed in polynomial time (for more details see~\cite{Schmid2013}). On the other hand, computing the locality number seems more difficult and it was left open in~\cite{DayEtAl2017} whether or not is is hard to compute. This gap was closed in~\cite{ICALP2019}  where it was shown that computing the locality number is $\npclass$-hard, but fixed-parameter tractable (if the locality number or $|\Sigma|$ is considered a parameter); in addition, approximation of the locality number has also been investigated in~\cite{ICALP2019} (note that these result will be discussed in more detail in Section~\ref{sec:cutwidth}).

\section{Faster Pattern Matching} \label{sec:fastMatch}

In this section we will overview some efficient matching algorithms developed for various classes of patterns, some defined already in the previous sections, and some defined via some other natural structural restrictions. Most of the result of this paper were shown in \cite{FeMaMeSc14_stacs,DayEtAl2017,DLT2018}.

\subsection{Patterns with Low Scope Coincidence Degree}\label{sec:algos}

We start with several definitions. 
%For any integer $k \geq 1$, a \emph{$k$-variable} pattern is a pattern $\alpha$ that satisfies $\left|\var(\alpha)\right| \leq k$ and a pattern $\beta$ with $|\{x \in \var(\beta) \mid |\beta|_x \geq 2\}| \leq k$ is a \emph{$k$-repeated-variable} pattern. 
%For every integer $k \geq 0$, $\var_k$ and $\repeatedvar_k$ denote the set of $k$-variable patterns and $k$-repeated-variable patterns, respectively. 
%Obviously, $\reg=\pat^{\mathsf{r}}_{\var \leq 0}$. 
The \emph{one-variable blocks} in a pattern are maximal contiguous blocks of occurrences of the same variable. A pattern $\alpha$ with $m$ one-variable blocks can be written as $\alpha=w_0\Pi_{i=1,m}(z_i^{k_i}w_i)$ with $z_i\in \var(\alpha)$
for $i\in \{1,2,\ldots,m\}$ and $z_i\neq z_{i+1}$, whenever $w_i=\varepsilon$ for $i\in\{1,2,\ldots,m-1\}$. The number of  one-variable blocks is a natural complexity measure that we will consider.\par

\begin{example}
The pattern $\alpha=x_1x_2x_2 \ta x_2x_2x_2x_3 \ta x_3x_2x_2x_3x_3$ has the following $7$ one-variable blocks: $x_1, x_2x_2, x_2x_2x_2, x_3, x_3, x_2 x_2, x_3 x_3$.
\end{example}

As discussed in the previous sections, prominent subclasses of patterns for which $\match$ can be solved in polynomial time are the classes of patterns with a bounded number of (repeated) variables ($\var_k$ and $\repeatedvar_k$), of regular patterns ($\regular$), of non-cross patterns ($\noncross$), and of patterns with a bounded scope coincidence degree ($\scd_{k}$). However, the known respective algorithms are rather poor considering their running times. For example, for $\var_{k}$, the matching problem can be solved in $\bigO(\frac{mn^{k-1}}{(k-1)!})$, where $m$ and $n$ are the lengths of the pattern and the word (see~\cite{iba:ano}). For patterns with a scope coincidence degree of at most $k$, an $\bigO(mn^{2(k+3)}(k+2)^2)$ time algorithm can be derived using the general matching technique described by Theorem~\ref{metaTheorem}, where $m$ and $n$ are the lengths of the pattern and the word, respectively, and the proof that the matching problem for non-cross patterns is in $\pclass$ (see~\cite{shi:pol2}) leads to an $\bigO(n^4)$-time algorithm. Hence, for all these classes, we consider the following refinement of the problem of showing that the matching problem for a class of patterns is in $\pclass$.

\begin{problem}
Let $K$ be a class of patterns for which the matching problem can be solved in polynomial time. Find an efficient algorithm that solves the matching problem for $K$.
\end{problem}

The main class of patters considered in the following is that of patterns with bounded scope coincidence degree, and its subclasses. 

If the scope coincidence degree is bounded by $1$, i.\,e., non-cross patterns, we can decide whether a pattern $\alpha$ having $m$ one-variable blocks matches a word $w$ of length $n$ in $\bigO(m n \log n)$ time. This result can be achieved via a general dynamic programming approach, which tries to match prefixes of the pattern $\alpha$ to the prefixes of the word $w$. This general approach is rather standard but the big gain is that it can be implemented efficiently by a a detailed combinatorial analysis of the possible matches between the one-variable blocks occurring in $\alpha$ and factors of $w$. For instance, if the shortest factor of $\alpha$ containing all occurrences of a variable $x$ starts with a one-variable block containing at least two occurrences the variable $x$, we can efficiently find the matches of this factor by exploiting a major result from~\cite{CrochemoreIPL81}, which states that the primitively rooted squares contained in a word of length $n$ can be listed optimally in $\bigO(n \log n)$. As each match for a factor starting with two occurrences of a variable starts with a primitively rooted square, the respective matches can be found efficiently. The result regarding primitively rooted squares can be extended to show that, given a word $w$ of length $n$ and a word $v$ with length shorter than $n$, the word $w$ contains $\bigO(n\log n)$ factors of the form $uvu$ with $uv$ primitive, and all these factors can be found optimally in $\bigO(n\log n)$ time. This allows us to find efficiently the matches for one-variable that the shortest factor of $\alpha$ which contains all occurrences of $x$ and starts with $x v x$, for all choices of a variable $x$  such that $v$ is a non-empty terminal string.

\begin{theorem}[\cite{FeMaMeSc14_stacs}]
The matching problem for $\noncross$ is solvable in $\bigO(m n \log n)$ time, where $w$ is the input word of length $n$ and $m$ is the number of one-variable blocks occurring in the pattern.
\end{theorem}

Two particular subclasses of non-cross patterns are of interest: the regular patterns $\regular$ and the one-variable patterns $\var_1$ (see also Figure~\ref{mainDiagramFigure}). 
%, i.e., patterns $\alpha$ such that $|\var(\alpha)|=1$. 
It is not hard to show that regular patterns can be matched in linear time $\bigO(|\alpha|+|w|)$, by iteratively using the Knuth-Morris-Pratt algorithm to identify greedily the terminal factors occurring in the pattern, in their orders of occurrences. All factors of a word $w$ that match a given regular pattern $\alpha$ can be detected in linear time too. 

More interesting is the case of one-variable patterns. The simplest example of one-variable patterns are the repetitions, i.e., patterns of the form $x^k$. Checking whether a word is a match for a pattern $x^k$ can be done in linear time. Moreover, a compact representation of all periodic factors of a word $w$ can be also obtained in linear time by identifying the (at most $|w|$) so-called runs inside $w$ \cite{BannaiIInenagaNakashimaTakedaTsuruta}. With this, a compact representation of occurrences of $x^k$ in $w$ can also be obtained in linear time. More complex one-variable patterns are the pseudo-repetitions (see \cite{GawrychowskiMMNT13,GawrychowskiManeaNowotka,GawrychowskiMMN19} and the references therein). These are patterns from $\{x,x^R\}^*$, where $x^R$ is a variable that is always substituted by the reverse image of the string substituting $x$. Checking whether a string matches a given pseudo-repetition can be done in linear time \cite{GawrychowskiMMNT13}. The following general result can be shown for one-variable patterns, see~\cite{KMN16-SPIRE}. Given a pattern $\alpha = v_1xv_2x\cdots v_{r-1}xv_r$ such that $x$ is a variable and $v_1,v_2,\ldots,v_r$ are terminal strings, a compact representation of all $P$ instances of $\alpha$ in the input string $w$ of length $n$ can be computed in $\bigO(rn)$ time, so that one can report those occurrences in $\bigO(P)$ time. The same result holds also for the case when some of the occurrences of $x$ in such a pattern are replaced by $x^R$. It is worth noting that using this algorithm to find the factors of a given word that match the shortest factor of $\alpha$ containing all occurrences of a variable $x$ inside a non-cross pattern in our approach for matching $\noncross$ does not lead to a faster matching algorithm in that case.

When considering general patterns with bounded scope coincidence degree, one can show, using a similar dynamic programming approach as in the case of non-cross patterns, that the matching problem for $\scd_k$ is solvable in $\bigO(\frac{m n^{2k}}{((k-1)!)^2})$ time, where $n$ is the length of the input word and $m$ is, again, the number of one-variable blocks occurring in the pattern. One should note that in this case it seems hard to use the combinatorial insights used for non-cross patterns (thus, the $\log n$ factor is replaced by an $n$ factor in the evaluation of the time complexity), but, still, this algorithm is significantly faster than the previously known solution.

\begin{theorem}[\cite{FeMaMeSc14_stacs}]
The matching problem for $\scd_k$ is solvable in $\bigO\left(\frac{mn^{2k}}{((k-1)!)^2}\right)$ time, where $w$ is the input word of length $n$ and $m$ is the number of one-variable blocks occurring in the pattern.
\end{theorem}

Next we consider the classes $\repeatedvar_k$. For the basic case of $k=1$, the matching problem can be solved in $\bigO(n^2)$ time, where $n$ is the length of the input word. 
The idea of this algorithm is to guess the length $\ell$ of the repeated variable $x$, and then to partition the suffix array of the input word into clusters, such that all suffixes in a cluster start with the same factor of length $\ell$. Essentially, in a match between the pattern and the word, where $x$ is mapped to a factor of length $\ell$, the positions where the factors matching $x$ occur in the input word belong to the same cluster. Using this idea, the desired complexity is then reached, again via dynamic programming. 

\begin{theorem}[\cite{FeMaMeSc14_stacs}]
\label{one_repeated}
The matching problem for $\repeatedvar_k$ is solvable in quadratic time.
%$\bigO(n^2)$ time, where $w$ is the input word of length $n$.
% \end{lemma} 
\end{theorem}

Further, one can use this result to show that the matching problem for the general class of patterns $\repeatedvar_k$ is solvable in $\bigO(\frac{n^{2k}}{((k-1)!)^2})$ time. This algorithm is better than the one that could have been obtained by using the fact that patterns with at most $k$ repeated variables have the scope coincidence degree bounded by $k+1$, and then directly applying our previous algorithm solving the matching problem for $\scd_{k+1}$.\par

\begin{theorem}[\cite{FeMaMeSc14_stacs}]
The matching problem for $\repeatedvar_k$ is solvable in $\bigO\left(\frac{n^{2k}}{((k-1)!)^2}\right)$ time, where $n$ is the length of the input word.
\end{theorem}

Note that the classes of non-cross patterns and of patterns with a bounded scope coincidence degree or with a bounded number of repeated variables are of special interest, since for them we can compute so-called descriptive patterns (see~\cite{ang:fin2,shi:pol2}) in polynomial time. A pattern $\alpha$ is \emph{descriptive} (with respect to, say, non-cross patterns) for a finite set $S$ of words if it can generate all words in $S$ and there exists no other non-cross pattern that describes the elements of $S$ in a better way. Computing a descriptive pattern, which is $\npclass$-complete in general, means to infer a pattern common to a finite set of words, with applications for inductive inference of pattern languages (see~\cite{ng:dev}). For example, our algorithm for computing non-cross patterns can be used in order to  obtain an algorithm that computes a descriptive non-cross pattern in time $\bigO(\sum_{w \in S} (m^2 |w| \log |w|))$, where $m$ is the length of a shortest word of $S$ (see~\cite{FeMaMeSc16_TCS} for details).\par

The algorithms, except the ones for the basic cases of regular and non-cross patterns and patterns with only one repeated variable, still have an exponential dependency on the number of repeated variables or the scope coincidence degree. 
Therefore, only for very low constant bounds on these parameters can these algorithms be considered efficient. Naturally, finding a polynomial time algorithm for which the degree of the polynomial does not depend on the number of repeated variables or on the scope coincidence degree would be desirable. However, by Remark \ref{no-fpt} such algorithms are very unlikely. 

Finally we recall a result regarding {\em gapped repeats and palindromes}. A gapped repeat (palindrome) is an instance of a terminal-free pattern $xyx$ (respectively, $xyx^R$). For $\alpha\geq 1$, an $\alpha$-gapped repeat in a word $w$ is a factor $uvu$ of $w$ such that $|uv|\leq \alpha |u|$; the two factors $u$ in such a repeat are called arms, while the factor $v$ is called gap. Such a repeat is called maximal if its arms cannot be extended simultaneously with the same symbol to the right or, respectively, to the left. In a sense, $\alpha$-gapped repeats are instances of the pattern $xyx$ where length constraints are imposed on the strings that substitute $x$ and $y$. In \cite{GawrychowskiIIK18} it was shown that the number of maximal $\alpha$-gapped repeats that may occur in a word is upper bounded by $18\alpha n$. Using this, an algorithm finding all the maximal $\alpha$-gapped repeats of a word in $\bigO(\alpha n)$ was defined; this result is optimal, in the worst case, as there are words that have $\Theta(\alpha n)$ maximal $\alpha$-gapped repeats. Comparable results were developed for the case of $\alpha$-gapped palindromes, i.e., factors $uvu^R$ with $|uv|\leq \alpha |u|$. On the one hand, these results were relevant as they provided optimal algorithms for the identification of $\alpha$-gapped repeats and palindromes, and closed an open problem from \cite{KK09,KolpakovPPK14} (see also \cite{GawrychowskiIIK18} and the references therein for more on gapped repeats and palindromes). On the other hand, they point towards the study of $\match$ for patterns with (linear) length constraints on the images of the variables.

\subsection{Patterns with Low Locality Number} \label{sec:local}

Intuitively, the notion of $k$-locality (already introduced in Section~\ref{sec:graphs}) involves marking the variables in the pattern in some 
arbitrary order until all the variables are marked. The pattern is $k$-local if this marking can be 
done while never creating more than $k$ marked blocks. Variables which only 
occur adjacent to those which are already marked can be marked ``for free'' -- without creating any 
new blocks, and thus a valid marking sequence allows a sort-of parsing of the pattern whilst 
maintaining a degree of closeness (locality) to the parts already parsed. The notion of $k$-locality was introduced and further analysed in \cite{DayEtAl2017}. With respect to pattern matching, the main result proven in that paper is the following:

\begin{theorem}[\cite{DayEtAl2017}]\label{match-k-loc}
$\match$ for $\loc_k$ can be decided in $O(mkn^{\max{(3k-1,2k+1)}})$ time, where $m$ is the length of the input pattern and $n$ is the length of the input word.
\end{theorem}
To solve the matching problem for $\loc_k$ we use the following idea. Using a simple dynamic programming approach we can show that, given a pattern $\beta \in (X\cup\Sigma)^*$ of length $m$, we can decide in $O(m^{2k}k)$ time whether $\beta\in \loc_k$, and if the answer is positive, we can produce in the same time a marking sequence witnessing that $\beta$ is $k$-local.
As such we can keep track of the marked factors in the pattern, while executing the marking according to the computed marking sequence. We also need now to keep track to which factors of the input word the marked factors correspond. Then we try to assign every new variable so that it fits nicely around the already matched factors. This is done efficiently using a data structure from \cite{KMN16-SPIRE}, mentioned also above: given a word $w$ and a one-variable pattern $\gamma$ (so, $|\var(\gamma)|=1$), one can produce a compact representation of all the $g$ factors of $w$ matching $\gamma$ in $O(|\gamma||w|)$ time; moreover, we can obtain all the $g$ factors of $w$ matching $\gamma$ in $O(|g|)$ time. This allows us to test efficiently which factors of $w$ match any of the one-variable blocks of $\beta$, and, ultimately, to assign a value to each variable. In comparison to the algorithm from~\cite{rei:patIaC} for patterns of bounded treewidth, which firstly constructs relational structures from $\alpha$ and $w$, and solves the homomorphism problem on these relational structures (see Section~\ref{sec:treewidth}), the above algorithm exploits directly the locality structure present in the patterns. The advantage of this more focussed approach is that it allows for a considerable improvement in the required time, reducing the exponent of $n$ from $4k+4$ to $3k-1$.

\section{Efficient Pattern Matching Beyond Bounded Treewidth} \label{sec:rep}

In \cite{DLT2018} the authors tried to identify classes of patterns that do not have bounded treewidth but can still be matched in polynomial time. The idea behind defining such classes was relatively simple: consider generalised repetitions of patterns.

One simple observation is that, if we can match patterns from a class $\mathcal{C}$ in polynomial time, then we can also match repetitions of these patterns in polynomial time: if we wish to check whether  $\alpha^k$ matches a word $w$, where $\alpha$ is chosen from the class $\mathcal{C}$ for which we can solve $\match$ efficiently, then we can firstly check whether $ w =v^k$ for some word $v$, and then check whether $\alpha$ matches $v$, so we can also match $\alpha^k$ efficiently. Moreover, it can be observed that most parameters that lead to efficiently matchable classes, e.\,g., the scope coincidence degree or locality, are defined independently from the terminal symbols, i.\,e., via the word obtained after removing all terminals, which shall be called \emph{skeleton} in the following (e.\,g., the skeleton of $x_1 \ta x_2 \tb \ta x_1 x_2 \tb$ is $x_1 x_2 x_1 x_2$). As a result, it is possible that a pattern, that is \emph{not} a repetition of any $\alpha \in \mathcal{C}$, has nevertheless a skeleton that is a repetition of a skeleton from $\mathcal{C}$. For example, $\ta x_1 (x_2)^3 x_3 \tb x_3 x_1 (x_2)^2 \tb x_2 \ta (x_3)^2$ is not a repetition of a non-cross pattern, but its skeleton $(x_1 (x_2)^3 (x_3)^2)^2$ is. In \cite{DLT2018} it is shown that, for some important classes ${\mathcal C}$ of patterns, including $\loc_k$ and $\scd_k$, for constant $k$, the polynomial time solvability of $\match$ does not only extend from $\mathcal{C}$ to exact repetitions, but also to such skeleton-repetitions, called \emph{$\mathcal{C}$-repetitions}.

\begin{theorem}[\cite{DLT2018}]
For $\mathcal{C} \in \{\noncross,\reg,\loc_k,\scd_k\}$, solving the matching problem for the class of $\mathcal{C}$-repetitions can be done in polynomial time.
\end{theorem} 

It is interesting to note that the general treewidth-based framework of polynomial time matching of patterns does not seem to cover a very simple and natural aspect: repetitions of the same pattern. More precisely, if $\mathcal{C}$ is one of the known efficiently matchable classes of patterns, then a repetition $\alpha^k$ for some $\alpha \in \mathcal{C}$ is usually not in $\mathcal{C}$ anymore. In fact, it can be shown that even for patterns $\alpha$ with bounded and very low treewidth, the treewidth of repetitions $\alpha^k$ can be unbounded. 

\begin{theorem}[\cite{DLT2018}]
Let ${\mathcal C}$ be a class of patterns that contains $\reg$. Then the class of ${\mathcal C}$-repetitions contains patterns with arbitrarily large treewidth.
\end{theorem}

In particular, the previous theorem holds for the class $\reg$ of regular patterns, arguably the simplest class allowing an unbounded number of variables (note that patterns with a constant number of variables can trivially be matched in polynomial-time). In the same paper it is shown that if the notion of repetition is relaxed further, by considering a setting where the order in which the variables appear is no longer constrained at all (i.e., considering {\em abelian repetitions} instead of repetitions), then the matching problem is $\npclass$-complete. This holds even in the minimal case when the number of repetitions is restricted to two, and that the pattern which is repeated is regular.

\section{From Locality to Graph Parameters} \label{sec:cutwidth}

Following the ideas of Section \ref{sec:hardness} we explore further the connection between string and graph parameters. The main idea behind such a connection is to reach it by ``flattening'' a graph into a sequential form, or by ``inflating'' a string into a graph, so that algorithmic techniques available for each one of these become applicable for the other one as well. 
In this section, following \cite{ICALP2019}, we are concerned with certain structural parameters (and the problems of computing them) for graphs and strings: the \emph{cutwidth} $\cutwidth(G)$ of a graph $G$ (i.\,e., the maximum number of ``stacked'' edges if the vertices of a graph are drawn on a straight line), the \emph{pathwidth} $\pathwidth(G)$ of a graph $G$ (i.\,e., the minimum width of a tree decomposition the tree structure of which is a path), and the \emph{locality number} $\loc(\alpha)$ of a string $\alpha$ (explained in more detail in Section~\ref{sec:graphs}). By $\cutwidthProb$, $\pathwidthProb$ and $\locProb$, we denote the corresponding natural decision problems (i.\,e., decide whether a given graph has a pathwidth/cutwidth, or a given string has a locality number of at most $k$, for given $k$) and with the prefix $\textsc{Min}$, we refer to the minimisation variants. The two former graph-parameters are very classical. Pathwidth is a simple (yet still hard to compute) subvariant of treewidth, which measures how much a graph resembles a path. The problems $\pathwidthProb$ and $\minPathwidthProb$ are intensively studied (in terms of exact, parameterised and approximation algorithms) and have numerous applications (see the surveys and textbook~\cite{Bodlaender1998,Kloks1994,Bodlaender1993}). $\cutwidthProb$ is the best known example of a whole class of so-called \emph{graph layout problems} (see the survey~\cite{surveyDiaz,Petit2011} for detailed information), which are studied since the 1970s and were originally motivated by questions of circuit layouts. \par
In comparison, the locality number seems a rather simple parameter directly defined on strings, but, however, it bounds the treewidth of the string (in the sense defined in Section~\ref{sec:treewidth}), and the corresponding marking sequences can be seen as instructions for a dynamic programming algorithm for matching the pattern. In this way, it resembles a bit to the way the pathwidth and treewidth of graphs are used in algorithmic settings. Moreover, compared to other ``tractability-parameters'' of strings, it seems to cover best the treewidth of a string, but it also cannot be efficiently computed compared to the other simpler parameters.\par 
Going more into detail, for $\locProb$, exact exponential-time algorithms are not hard to be devised \cite{DayEtAl2017} but whether it can be solved in polynomial-time, or whether it is at least fixed-parameter tractable was left open in the paper where this measure was introduced. On the other hand, $\pathwidthProb$ and $\cutwidthProb$ are known $\npclass$-complete problems, fixed-parameter tractable with respect to parameter $\pathwidth(G)$ or $\cutwidth(G)$, respectively (even with ``linear'' fpt-algorithms with running-time $g(k) \bigO(n)$~\cite{Bodlaender1996,Bodlaender2012,ThilikosSB05}). With respect to approximation, their minimisation variants have received a lot of attention, mainly because they yield (like many other graph parameters) general algorithmic approaches for numerous graph problems, i.\,e., a good linear arrangement or path-decomposition can often be used to design a dynamic programming (or even divide and conquer) algorithm for other problems. The best known approximation algorithms for the problems $\minPathwidthProb$ and $\minCutwidthProb$ (with approximations ratios of $\bigO(\sqrt{\log(\opt)} \log(n))$ and $\bigO(\log^2(n))$, respectively) follow from approximations of vertex separators (see~\cite{FeigeEtAl2008}) and edge separators (see~\cite{Leighton1999}), respectively. \par
 There are two natural approaches to represent a word $\alpha$ over alphabet $\Sigma$ as a graph $G_{\alpha} = (V_{\alpha}, E_{\alpha})$: (1) $V_{\alpha} = \{1, 2, \ldots, |\alpha|\}$ and the edges are somehow used to represent the actual symbols (note that this is the case for the standard graph representation of patterns defined in Section~\ref{sec:treewidth}), or (2) $V_{\alpha} = \Sigma$ and the edges are somehow used to represent the positions of $\alpha$. A reduction of type (2) can be defined such that $|E_{\alpha}| = \bigO(|\alpha|)$ and $\cutwidth(G_{\alpha}) = 2\loc(\alpha)$, and a reduction of type (1) can be defined such that $|E_{\alpha}| = \bigO(|\alpha|^2)$ and $\loc(\alpha) \leq \pathwidth(G_{\alpha}) \leq 2\loc(\alpha)$. Since these reductions are parameterised reductions and also allow to transfer approximation results, one may conclude that $\locProb$ is fixed-parameter tractable if parameterised by $|\Sigma|$ (note that for parameter $|\Sigma|$ a simple, but less efficient fpt-algorithm is trivially obtained by simply enumerating all marking sequences) or by the locality number, and also that there is a polynomial-time $\bigO(\sqrt{\log(\opt)} \log(n))$-approximation algorithm for $\minlocProb$. \par
In addition, one can represent an arbitrary multi-graph $G = (V, E)$ by a word $\alpha_G$ over alphabet $V$ with $|\alpha_G|=|E|$ and $\cutwidth(G) = \loc(\alpha)$. This describes a Turing-reduction from $\cutwidthProb$ to $\locProb$ which also allows to transfer approximation results between the minimisation variants.
% from $\minCutwidthProb$ to $\minlocProb$. 
As a result, $\locProb$ is $\npclass$-complete. Finally, by plugging together the reductions from $\minCutwidthProb$ to $\minlocProb$ and from $\minlocProb$ to $\minPathwidthProb$, one obtains a reduction which transfers approximation results from $\minPathwidthProb$ to $\minCutwidthProb$, which yields an $\bigO(\sqrt{\log(\opt)} \log(n))$-approximation algorithm for $\minCutwidthProb$. This result from \cite{ICALP2019} improved, for the first time since 1999, the best approximation for $\cutwidthProb$ from~\cite{Leighton1999}. Interestingly, this improvement appeared as a side-product of relating string-parameters with graph-parameters.
% by this very unusual approach to graph problems via strings. \par
\begin{theorem}[\cite{ICALP2019}]\label{thm:approxLoc1}
There is an $\bigO(\sqrt{\log(\opt)} \log(h))$-approximation algorithm (running in polynomial time) for $\minCutwidthProb$ on multigraphs with $h$ edges.
In particular, this yields an $\bigO(\sqrt{\log(\opt)} \log(n))$-approximation algorithm for $\minCutwidthProb$ for graphs.
\end{theorem}

Moreover, this approach allows also for establishing a direct connection between cutwidth and pathwidth, which preserves the good algorithmic properties, and has not yet been reported in the literature so far. This is rather surprising, since $\cutwidthProb$ and $\pathwidthProb$ have been jointly investigated in the context of exact and approximation algorithms, especially in terms of balanced vertex and edge separators.
We think that a reason for overlooking this connection might be that it is less obvious on the graph level and becomes more apparent if linked via the string parameter of locality, emphasising, as such, the value of such mixed approaches.

\section{Extensions}

\subsection{Injectivity}

In our setting, the substitutions that map variables to words are \emph{not} required to be injective, i.\,e., different variables can be mapped to the same word. However, the requirement of injectivity is natural in some contexts. For example, in the pattern matching community, the first mentioning of pattern matching with variables concerns the case where variables have to be substituted by single symbols and in an injective way. More precisely, this \emph{parameterised pattern matching} was introduced in~\cite{bak:par} to formalise the problem of detecting code clones (i.\,e., we want to find code segments that are created by copying some code blocks and renaming program variables (this renaming will be injective, since otherwise the semantic of the code might change)). More generally speaking, the injectivity condition is appropriate whenever we know a priori that different variables should always refer to different words (e.\,g., when matching the pattern $$x_1\:\textsf{name:}\:y\:;\:\textsf{address:}\:z\:;\:x_2\:\textsf{name:}\:y\:;\:\textsf{address:}\:z\:x_3$$ in order to check whether there is a repetition of some name-address data tuple, then it is likely that we can assume injectivity). \par
Depending on the actual variant, the injectivity condition can make the matching problem harder or easier. In~\cite{FeMaMeSc14_stacs}, it is shown that it is $\npclass$-hard to decide for a given word $w$ and an integer $k$ whether $w$ can be factorised into at least $k$ \emph{pairwise different} factors. This immediately implies that the injectivity condition makes the matching problem $\npclass$-hard even for the ``trivial'' pattern class $\{x_1 x_2 \ldots x_n \mid n \geq 1\}$ (note that this is even a subset of the class $\regular$ of regular patterns). On the other hand, if we have an upper bound on $|\Sigma|$ and $\max\{|h(x)| \mid x \in X\}$ (recall that this case is still $\npclass$-hard even for bounds $2$ and $1$, respectively; see Theorem~\ref{hardnessExampleTheorem}) then also the total number of possible substitution words is bounded; thus, the injectivity condition bounds the total number of variables and therefore the matching problem becomes tractable (see~\cite{FernauSchmid2015}). A similar observation can be made with respect to fixed-parameter tractability if we parameterise by $|\Sigma|$ and $\max\{|h(x)| \mid x \in X\}$ (see~\cite{FernauEtAl2016}).

%On the other hand, if we have an upper bound on $|\Sigma|$ and $\max\{|h(x)| \mid x \in X\}$ (recall that this case is still $\npclass$-hard even for bounds $2$ and $1$, respectively; see Theorem~\ref{hardnessExampleTheorem}) then the injectivity conditions leads to tractability, since then also the total number of variables is bounded (see~\cite{FernauSchmid2015}). A similar observation can be made with respect to fixed-parameter tractability if we parameterise by $|\Sigma|$ and $\max\{|h(x)| \mid x \in X\}$ (see~\cite{FernauEtAl2016}).

\subsection{Word Equations}

A {\em word equation} is an equality $\alpha = \beta$, where $\alpha$ and $\beta$ are patterns with variables, e.\,g., $\alpha =x_1 \ta \tb x_2$ and $\beta = \ta x_1 x_2 \tb$ define the equation $x_1 \ta \tb x_2 = \ta x_1 x_2 \tb$. A \emph{solution} to an equation $\alpha = \beta$ is a substitution $h : (\var(\alpha) \cup \var(\beta)) \rightarrow \Sigma^*$ (in the sense defined in Section~\ref{sec:basicDef}) that satisfies $h(\alpha) = h(\beta)$. For the example equation from above, the solutions are the substitutions $h$ with $h(x_1) = \ta^k$, for $k\geq 0$, and $h(x_2) = \tb^\ell$, for $\ell \geq 0$. \par
The study of word equations (or the existential theory of equations over free monoids) is an important topic found at the intersection of algebra and computer science, with significant connections to, e.g., combinatorial group or monoid theory~\cite{Lyndon77,Lyndon60,DiekertW16}, unification~\cite{sch:wor,Jaffar90,Jez14}), and, more recently, data base theory \cite{FreydenbergerHolldack18,Freydenberger2018}.\par
The central computational problem for word equations is the satisfiability problem, i.\,e., the problem of deciding whether a given word equation $\alpha=\beta$ has a solution or not. In this regard, the matching problem for patterns with variables describes just the special case of the satisfiability problem for word equations where one side of the equation is a terminal word, e.\,g., $x_1 \ta \tb x_1 x_2 \tc x_2 x_1 = \tb \ta \tb \tb \ta \tc \ta \tb$ is an instance of the matching problem already mentioned in the introduction, phrased as a word equation. Consequently, the satisfiability problem is intractable, even for very strongly restricted cases (see Theorems~\ref{hardnessExampleTheorem}~and~\ref{FPThardnessExampleTheorem}). Also note that it has been shown in~\cite{DieRob99} that the solvability problem remains $\npclass$-hard if every variable has at most two occurrences in $\alpha\beta$ (called \emph{quadratic} equations), but the proof of~\cite{DieRob99} actually talks about the matching problem for patterns with at most two occurrences per variable. \par
While the matching problem for patterns with variables is trivially decidable, it is not at all obvious how to solve the satisfiability problem for word equations. In fact, the question whether it is decidable was initially approached with the expectation that it will be answered in the negative. It was, however, shown to be decidable by Makanin~\cite{mak:the} (see Chapter~$12$ of \cite{lot:alg} for a survey). Later it was shown that the satisfiability problem is in $\pspaceclass$ by Plandowski~\cite{Pla2006}; a new proof of this result was obtained in~\cite{Jez2016b}, based on a new simple technique called recompression. There are also cases when the satisfiability problem is tractable. For instance, word equations with only one variable can be solved in linear time in the size of the equation, see~\cite{Jez2016}; equations with two variables can be solved in time $\bigO(|\alpha\beta|^5)$, see~\cite{DabPla2004}.\par
Given the fact that there are many structural restrictions of patterns that yield tractability (with respect to the matching problem, see Section~\ref{sec:Classes}), the question naturally arises how the complexity of the satisfiability problem for word equations (which are essentially equations of patterns) behaves if these restrictions are applied to word equations. More precisely, while each class of patterns with $\npclass$-hard matching problem yields a class of word equations with $\npclass$-hard satisfiability problem, the hardness of the satisfiability problem for equations with sides in some efficiently matchable class of patterns is no longer immediate. An investigation of that question was initiated in~\cite{ManSchNow16}, where the following results were obtained. Firstly, the satisfiability problem for non-cross word equations (i.\,e., word equations for which both sides are non-cross) remains $\npclass$-hard. In particular, solving non-cross equations $\alpha=\beta$ where each variable occurs at most three times, at most twice in $\alpha$ and exactly once in $\beta$, is $\npclass$-hard (note that this constitutes the first $\npclass$-hardness result for word equations that is not a direct conclusion from a hardness result for the matching problem). Secondly, the satisfiability of one-repeated variable equations (i.\,e., at most one variable occurs more than once in $\alpha\beta$, but arbitrarily many other variables occur only once) having at least one non-repeated variable on each side, was shown to be trivially in~$\pclass$.\par
In \cite{MFCS2017}, it is shown that it is (still) $\npclass$-hard to solve regular ordered word equations. More precisely, these are word equations where each side is a regular pattern and the order of the variables in both sides is the same (it is, however, possible that some variables only occur on one side of the equation), e.\,g., $x_1 \ta x_2 \tb \ta x_3 x_4 = \tb x_1 x_3 \ta \ta x_4$ is a regular ordered word equation. They are particular cases of both quadratic equations and non-cross equations, so the reductions showing the hardness of solving these more general equations do not carry over. In particular, note that the class of regular patterns is arguably the most simple class of patterns in terms of their matching complexity (see Section~\ref{sec:algos}).\par
The respective hardness reduction relied on some deep word-combinatorics ideas. As a first step, a reachability problem for a certain type of (regulated) string rewriting systems was introduced, and showed it is $\npclass$-complete. This was achieved via a reduction from the problem \textsc{3-Partition}~\cite{gar:com}, which is strongly $\npclass$-complete. Then it was shown that this reachability problem can be reduced to the satisfiability of regular-ordered word equations; in this reduction the applications of the rewriting rules of the system were encoded into the periods of the words assigned to the variables in a solution to the equation. The main technicality was to make sure to only use one occurrence of each variable per side, and moreover to even have the variables in the same order in both sides. This result exhibits the arguably structurally-simplest class of word equations for which the satisfiability problem is $\npclass$-hard.\par
The main open problem in the area of word equations remains, even for simple subclasses such as regular equations or quadratic equations, to show that the satisfiability problem of word equations of the respective types is in $\npclass$ (note that this was already explicitly posed as an open question for the class of quadratic word equations in~\cite{DieRob99}).

\section{Conclusions}

In this work we tried to survey several results related to the problem of matching patterns with variables, that seem important to us. While this work is clearly not exhaustive, it is aimed to offer a basic understanding of the problems and state of the art in this area. 

From an algorithmic point of view, the results we covered provide a wide variety of classes of patterns with variables, for which $\match$ can be efficiently solved. Moreover, as explained in Section \ref{struct}, it is usually easy to check whether a pattern belongs to one of these classes. So, putting it all together, one could use the following approach when trying to match a pattern, rather than just using an exponential time algorithm (based, e.g., on general SMT-solvers, or on the theory of string solving \cite{barrett2011,zheng2017}). First, check if the pattern belongs to one of the classes for which efficient matching algorithms are known and, then, use this algorithm; only use a general algorithm when no customised one can be applied. Identifying more natural pattern classes for which $\match$ can be solved efficiently appears, as such, as a rather useful task. Following the practically motivated challenges that arise from the area of string solving, one could also try to find efficient matching algorithms for various classes of patterns, enhanced with various constraints: regular constraints, length constraints, etc. 

As an important part of this survey deals with polynomial time algorithms, it is natural to also ask whether they are optimal or not. This kind of questions are the focus of the area of fine-grained complexity (see, e.g., the survey \cite{bringmann:LIPIcs:2019:10243} and the citations therein). It would be interesting to see, using tools from this area, whether one can show lower bounds for the $\match$ problems for different classes of patterns.

In the light of the results from \cite{ICALP2019}, it seems that exploring the connections between string parameters and parameters for other classes of objects could lead to some interesting results in both worlds. So, it also seems like an interesting challenge to explore what the structural parameters of strings that we explored here (and maybe some other new ones) mean when various other types of data are represented as strings, and what consequences can be derived from such a representation.

Finally, the area of word equations abounds with open problems. As mentioned, it is not even clear whether the satisfiability of regular or quadratic equations is in $\npclass$. So even if we restrict to equations with structurally simple left and right hand sides, the complexity of solving equations is not known. Such problems become even more involved when we consider equations with various types of constraints (e.g., length or regular). For instance, the decidability of general word equations with length constraints is a long standing open problem, but it is already an interesting open question for simpler cases (once again: regular or quadratic equations); see, e.g., \cite{HalfonSZ17,DayGHMN18,LinM18}, and the references therein. It seems interesting to us whether some of the ideas used in matching patterns can be transferred to solving (simplified) word equations, with or without constraints.

 \bibliographystyle{plainurl}
 \bibliography{bibFile}

\begin{thebibliography}{10}

\bibitem{ami:gen}
Amihood Amir and Igor Nor.
\newblock Generalized function matching.
\newblock {\em Journal of Discrete Algorithms}, 5:514--523, 2007.

\bibitem{ang:fin2}
Dana Angluin.
\newblock Finding patterns common to a set of strings.
\newblock {\em Journal of Computer and System Sciences}, 21:46--62, 1980.

\bibitem{bak:par}
Brenda~S. Baker.
\newblock Parameterized pattern matching: Algorithms and applications.
\newblock {\em Journal of Computer and System Sciences}, 52:28--42, 1996.

\bibitem{BannaiIInenagaNakashimaTakedaTsuruta}
Hideo Bannai, Tomohiro I, Shunsuke Inenaga, Yuto Nakashima, Masayuki Takeda,
  and Kazuya Tsuruta.
\newblock The "runs" theorem.
\newblock {\em {SIAM} J. Comput.}, 46(5):1501--1514, 2017.

\bibitem{bar:exp}
Pablo Barcel\'{o}, Leonid Libkin, Anthony~W. Lin, and Peter~T. Wood.
\newblock Expressive languages for path queries over graph-structured data.
\newblock {\em ACM Transactions on Database Systems}, 37, 2012.

\bibitem{barrett2011}
Clark Barrett, Christopher~L Conway, Morgan Deters, Liana Hadarean, Dejan
  Jovanovi{\'c}, Tim King, Andrew Reynolds, and Cesare Tinelli.
\newblock {CVC4}.
\newblock In {\em International Conference on Computer Aided Verification},
  pages 171--177. Springer, 2011.

\bibitem{Bodlaender1993}
Hans~L. Bodlaender.
\newblock A tourist guide through treewidth.
\newblock {\em Acta Cybern.}, 11(1-2):1--21, 1993.

\bibitem{Bodlaender1996}
Hans~L. Bodlaender.
\newblock A linear-time algorithm for finding tree-decompositions of small
  treewidth.
\newblock {\em {SIAM} J. Comput.}, 25(5):1305--1317, 1996.
\newblock \href {http://dx.doi.org/10.1137/s0097539793251219}
  {\path{doi:10.1137/s0097539793251219}}.

\bibitem{Bodlaender1998}
Hans~L. Bodlaender.
\newblock A partial \emph{k}-arboretum of graphs with bounded treewidth.
\newblock {\em Theor. Comput. Sci.}, 209(1-2):1--45, 1998.
\newblock \href {http://dx.doi.org/10.1016/S0304-3975(97)00228-4}
  {\path{doi:10.1016/S0304-3975(97)00228-4}}.

\bibitem{Bodlaender2012}
Hans~L. Bodlaender.
\newblock Fixed-parameter tractability of treewidth and pathwidth.
\newblock In Hans~L. Bodlaender, Rod Downey, Fedor~V. Fomin, and D{\'{a}}niel
  Marx, editors, {\em The Multivariate Algorithmic Revolution and Beyond},
  volume 7370 of {\em LNCS}, pages 196--227, 2012.

\bibitem{bringmann:LIPIcs:2019:10243}
Karl Bringmann.
\newblock {Fine-Grained Complexity Theory (Tutorial)}.
\newblock In Rolf Niedermeier and Christophe Paul, editors, {\em 36th
  International Symposium on Theoretical Aspects of Computer Science (STACS
  2019)}, volume 126 of {\em Leibniz International Proceedings in Informatics
  (LIPIcs)}, pages 4:1--4:7, Dagstuhl, Germany, 2019. Schloss
  Dagstuhl--Leibniz-Zentrum fuer Informatik.
\newblock URL: \url{http://drops.dagstuhl.de/opus/volltexte/2019/10243}, \href
  {http://dx.doi.org/10.4230/LIPIcs.STACS.2019.4}
  {\path{doi:10.4230/LIPIcs.STACS.2019.4}}.

\bibitem{cam:afo}
Cezar C\^ampeanu, Kai Salomaa, and Sheng Yu.
\newblock A formal study of practical regular expressions.
\newblock {\em International Journal of Foundations of Computer Science},
  14:1007--1018, 2003.

\bibitem{ICALP2019}
Katrin Casel, Joel~D. Day, Pamela Fleischmann, Tomasz Kociumaka, Florin Manea,
  and Markus~L. Schmid.
\newblock Graph and string parameters: Connections between pathwidth, cutwidth
  and the locality number.
\newblock {\em CoRR, to appear in Proc. ICALP 2019,}, abs/1902.10983, 2019.
\newblock URL: \url{http://arxiv.org/abs/1902.10983}, \href
  {http://arxiv.org/abs/1902.10983} {\path{arXiv:1902.10983}}.

\bibitem{CrochemoreIPL81}
Maxime Crochemore.
\newblock An optimal algorithm for computing the repetitions in a word.
\newblock {\em Information Processing Letters}, 12(5):244--250, 1981.

\bibitem{DayEtAl2017}
Joel~D. Day, Pamela Fleischmann, Florin Manea, and Dirk Nowotka.
\newblock Local patterns.
\newblock In {\em 37th {IARCS} Annual Conference on Foundations of Software
  Technology and Theoretical Computer Science, {FSTTCS} 2017}, pages
  24:1--24:14, 2017.

\bibitem{DLT2018}
Joel~D. Day, Pamela Fleischmann, Florin Manea, Dirk Nowotka, and Markus~L.
  Schmid.
\newblock On matching generalised repetitive patterns.
\newblock In {\em Proc. Developments in Language Theory - 22nd International
  Conference, {DLT} 2018}, volume 11088 of {\em Lecture Notes in Computer
  Science}, pages 269--281. Springer, 2018.

\bibitem{DayGHMN18}
Joel~D. Day, Vijay Ganesh, Paul He, Florin Manea, and Dirk Nowotka.
\newblock The satisfiability of word equations: Decidable and undecidable
  theories.
\newblock In {\em Proc. 12th International Conference Reachability Problems,
  {RP} 2018}, volume 11123 of {\em Lecture Notes in Computer Science}, pages
  15--29. Springer, 2018.

\bibitem{MFCS2017}
Joel~D. Day, Florin Manea, and Dirk Nowotka.
\newblock The hardness of solving simple word equations.
\newblock In {\em Proc. {MFCS} 2017}, volume~83 of {\em LIPIcs}, pages
  18:1--18:14, 2017.

\bibitem{DabPla2004}
R.~D\c{a}browski and W.~Plandowski.
\newblock Solving two-variable word equations.
\newblock In {\em Proc. 31th International Colloquium on Automata, Languages
  and Programming, ICALP 2004}, volume 3142 of {\em Lecture Notes in Computer
  Science}, pages 408--419, 2004.

\bibitem{surveyDiaz}
Josep D\'{\i}az, Jordi Petit, and Maria Serna.
\newblock A survey of graph layout problems.
\newblock {\em ACM Comput. Surv.}, 34(3):313--356, September 2002.
\newblock \href {http://dx.doi.org/10.1145/568522.568523}
  {\path{doi:10.1145/568522.568523}}.

\bibitem{DiekertW16}
V.~Diekert, A.~Jez, and M.~Kufleitner.
\newblock Solutions of word equations over partially commutative structures.
\newblock In {\em Proc. 43rd International Colloquium on Automata, Languages,
  and Programming, {ICALP} 2016}, volume~55 of {\em Leibniz International
  Proceedings in Informatics (LIPIcs)}, pages 127:1--127:14, 2016.

\bibitem{DieRob99}
V.~Diekert and J.~M. Robson.
\newblock On quadratic word equations.
\newblock In {\em Proc. 16th Annual Symposium on Theoretical Aspects of
  Computer Science, STACS 1999}, volume 1563 of {\em Lecture Notes in Computer
  Science}, pages 217--226, 1999.

\bibitem{DowneyFellows2013}
Rodney~G. Downey and Michael~R. Fellows.
\newblock {\em Fundamentals of Parameterized Complexity}.
\newblock Texts in Computer Science. Springer, 2013.

\bibitem{erl:lea}
Thomas Erlebach, Peter Rossmanith, Hans Stadtherr, Angelika Steger, and Thomas
  Zeugmann.
\newblock Learning one-variable pattern languages very efficiently on average,
  in parallel, and by asking queries.
\newblock {\em Theoretical Computer Science}, 261:119--156, 2001.

\bibitem{FeigeEtAl2008}
Uriel Feige, MohammadTaghi HajiAghayi, and James~R. Lee.
\newblock Improved approximation algorithms for minimum weight vertex
  separators.
\newblock {\em {SIAM} J. Comput.}, 38(2):629--657, 2008.
\newblock \href {http://dx.doi.org/10.1137/05064299x}
  {\path{doi:10.1137/05064299x}}.

\bibitem{FeMaMeSc14_stacs}
Henning Fernau, Florin Manea, Robert Mercas, and Markus~L. Schmid.
\newblock Pattern matching with variables: Fast algorithms and new hardness
  results.
\newblock In {\em 32nd International Symposium on Theoretical Aspects of
  Computer Science, {STACS} 2015}, pages 302--315, 2015.

\bibitem{FeMaMeSc16_TCS}
Henning Fernau, Florin Manea, Robert Mercas, and Markus~L. Schmid.
\newblock Revisiting shinohara's algorithm for computing descriptive patterns.
\newblock {\em Theoretical Computer Science}, 733:44--54, 2018.

\bibitem{FernauSchmid2015}
Henning Fernau and Markus~L. Schmid.
\newblock Pattern matching with variables: {A} multivariate complexity
  analysis.
\newblock {\em Inf. Comput.}, 242:287--305, 2015.

\bibitem{FernauEtAl2016}
Henning Fernau, Markus~L. Schmid, and Yngve Villanger.
\newblock On the parameterised complexity of string morphism problems.
\newblock {\em Theory Comput. Syst.}, 59(1):24--51, 2016.

\bibitem{FlumGrohe06}
J{\"{o}}rg Flum and Martin Grohe.
\newblock {\em Parameterized Complexity Theory}.
\newblock Texts in Theoretical Computer Science. An {EATCS} Series. Springer,
  2006.

\bibitem{Fre2013}
Dominik~D. Freydenberger.
\newblock Extended regular expressions: Succinctness and decidability.
\newblock {\em Theory of Computing Systems}, 53:159--193, 2013.

\bibitem{Freydenberger2018}
Dominik~D. Freydenberger.
\newblock A logic for document spanners.
\newblock {\em Theory of Computing Systems}, Sep 2018.
\newblock URL: \url{https://doi.org/10.1007/s00224-018-9874-1}.

\bibitem{FreydenbergerHolldack18}
Dominik~D. Freydenberger and Mario Holldack.
\newblock Document spanners: From expressive power to decision problems.
\newblock {\em Theory of Computing Systems}, 62(4):854--898, 2018.

\bibitem{FreydenbergerReidenbach2010b}
Dominik~D. Freydenberger and Daniel Reidenbach.
\newblock Bad news on decision problems for patterns.
\newblock {\em Inf. Comput.}, 208(1):83--96, 2010.

\bibitem{FreydenbergerReidenbach2010}
Dominik~D. Freydenberger and Daniel Reidenbach.
\newblock Existence and nonexistence of descriptive patterns.
\newblock {\em Theor. Comput. Sci.}, 411(34-36):3274--3286, 2010.

\bibitem{FreydenbergerReidenbach2013}
Dominik~D. Freydenberger and Daniel Reidenbach.
\newblock Inferring descriptive generalisations of formal languages.
\newblock {\em J. Comput. Syst. Sci.}, 79(5):622--639, 2013.

\bibitem{fri:mas}
Jeffrey~E.~F. Friedl.
\newblock {\em Mastering Regular Expressions}.
\newblock O'Reilly, Sebastopol, CA, third edition, 2006.

\bibitem{gar:com}
Michael~R. Garey and David~S. Johnson.
\newblock {\em Computers and Intractability: A Guide to the Theory of
  NP-Completeness}.
\newblock W.~H. Freeman \& Co., New York, NY, USA, 1979.

\bibitem{GawrychowskiManeaNowotka}
P.~Gawrychowski, F.~Manea, and D.~Nowotka.
\newblock Testing generalised freeness of words.
\newblock In {\em STACS 2014}, volume~25 of {\em LIPIcs}, pages 337--349.
  Schloss Dagstuhl--Leibniz-Zentrum fuer Informatik, 2014.

\bibitem{GawrychowskiIIK18}
Pawel Gawrychowski, Tomohiro I, Shunsuke Inenaga, Dominik K{\"{o}}ppl, and
  Florin Manea.
\newblock Tighter bounds and optimal algorithms for all maximal
  {\(\alpha\)}-gapped repeats and palindromes - finding all maximal
  {\(\alpha\)}-gapped repeats and palindromes in optimal worst case time on
  integer alphabets.
\newblock {\em Theory Comput. Syst.}, 62(1):162--191, 2018.

\bibitem{GawrychowskiMMN19}
Pawel Gawrychowski, Florin Manea, Robert Mercas, and Dirk Nowotka.
\newblock Hide and seek with repetitions.
\newblock {\em J. Comput. Syst. Sci.}, 101:42--67, 2019.
\newblock URL: \url{https://doi.org/10.1016/j.jcss.2018.10.004}, \href
  {http://dx.doi.org/10.1016/j.jcss.2018.10.004}
  {\path{doi:10.1016/j.jcss.2018.10.004}}.

\bibitem{GawrychowskiMMNT13}
Pawel Gawrychowski, Florin Manea, Robert Mercas, Dirk Nowotka, and Catalin
  Tiseanu.
\newblock Finding pseudo-repetitions.
\newblock In {\em 30th International Symposium on Theoretical Aspects of
  Computer Science, {STACS} 2013, February 27 - March 2, 2013, Kiel, Germany},
  volume~20 of {\em LIPIcs}, pages 257--268, 2013.

\bibitem{gei:lea}
M.~Geilke and S.~Zilles.
\newblock Learning relational patterns.
\newblock In {\em Proc. 22nd International Conference on Algorithmic Learning
  Theory, ALT 2011}, volume 6925 of {\em Lecture Notes in Computer Science},
  pages 84--98, 2011.

\bibitem{HalfonSZ17}
Simon Halfon, Philippe Schnoebelen, and Georg Zetzsche.
\newblock Decidability, complexity, and expressiveness of first-order logic
  over the subword ordering.
\newblock In {\em Proc. 32nd Annual {ACM/IEEE} Symposium on Logic in Computer
  Science, {LICS} 2017}, pages 1--12. {IEEE} Computer Society, 2017.

\bibitem{iba:ano}
Oscar~H. Ibarra, Ting-Chuen Pong, and Stephen~M. Sohn.
\newblock A note on parsing pattern languages.
\newblock {\em Pattern Recognition Letters}, 16:179--182, 1995.

\bibitem{Jaffar90}
J.~Jaffar.
\newblock Minimal and complete word unification.
\newblock {\em Journal of the {ACM}}, 37(1):47--85, 1990.

\bibitem{Jez14}
A.~Jez.
\newblock Context unification is in {PSPACE}.
\newblock In {\em Proc. 41st International Colloquium on Automata, Languages,
  and Programming, {ICALP} 2014}, volume 8573 of {\em Lecture Notes in Computer
  Science}, pages 244--255. Springer, 2014.

\bibitem{Jez2016}
A.~Je\.{z}.
\newblock One-variable word equations in linear time.
\newblock {\em Algorithmica}, 74:1--48, 2016.

\bibitem{Jez2016b}
A.~Je\.{z}.
\newblock Recompression: A simple and powerful technique for word equations.
\newblock {\em Journal of the ACM}, 63, 2016.

\bibitem{JiangEtAl1995}
Tao Jiang, Arto Salomaa, Kai Salomaa, and Sheng Yu.
\newblock Decision problems for patterns.
\newblock {\em J. Comput. Syst. Sci.}, 50(1):53--63, 1995.

\bibitem{kar:the}
Juhani Karhum\"aki, Wojciech Plandowski, and Filippo Mignosi.
\newblock The expressibility of languages and relations by word equations.
\newblock {\em Journal of the ACM}, 47:483--505, 2000.

\bibitem{KaSaBu06}
Juha K\"{a}rkk\"{a}inen, Peter Sanders, and Stefan Burkhardt.
\newblock Linear work suffix array construction.
\newblock {\em Journal of the ACM}, 53:918--936, 2006.

\bibitem{kea:apo}
Michael~J. Kearns and Leonard Pitt.
\newblock A polynomial-time algorithm for learning \emph{k-}variable pattern
  languages from examples.
\newblock In {\em Proceedings of the Second Annual Workshop on Computational
  Learning Theory, {COLT} 1989, Santa Cruz, CA, USA, July 31 - August 2,
  1989.}, pages 57--71, 1989.

\bibitem{Kloks1994}
Ton Kloks.
\newblock {\em Treewidth, Computations and Approximations}, volume 842 of {\em
  Lecture Notes in Computer Science}.
\newblock Springer, 1994.
\newblock \href {http://dx.doi.org/10.1007/BFb0045375}
  {\path{doi:10.1007/BFb0045375}}.

\bibitem{KK09}
Roman Kolpakov and Gregory Kucherov.
\newblock Searching for gapped palindromes.
\newblock {\em Theor. Comput. Sci.}, 410(51):5365--5373, 2009.

\bibitem{KolpakovPPK14}
Roman Kolpakov, Mikhail Podolskiy, Mikhail Posypkin, and Nickolay Khrapov.
\newblock Searching of gapped repeats and subrepetitions in a word.
\newblock In {\em Proc. 25th Annual Symposium Combinatorial Pattern Matching,
  {CPM} 2014}, volume 8486 of {\em Lecture Notes in Computer Science}, pages
  212--221. Springer, 2014.

\bibitem{KMN16-SPIRE}
Dmitry Kosolobov, Florin Manea, and Dirk Nowotka.
\newblock Detecting one-variable patterns.
\newblock In {\em String Processing and Information Retrieval - 24th
  International Symposium, {SPIRE} 2017, Palermo, Italy, September 26-29, 2017,
  Proceedings}, pages 254--270, 2017.

\bibitem{Leighton1999}
Tom Leighton and Satish Rao.
\newblock Multicommodity max-flow min-cut theorems and their use in designing
  approximation algorithms.
\newblock {\em J. ACM}, 46(6):787--832, November 1999.
\newblock \href {http://dx.doi.org/10.1145/331524.331526}
  {\path{doi:10.1145/331524.331526}}.

\bibitem{LinM18}
Anthony~W. Lin and Rupak Majumdar.
\newblock Quadratic word equations with length constraints, counter systems,
  and presburger arithmetic with divisibility.
\newblock In {\em Proc. 16th International Symposium Automated Technology for
  Verification and Analysis, {ATVA} 2018}, volume 11138 of {\em Lecture Notes
  in Computer Science}, pages 352--369. Springer, 2018.

\bibitem{Loth97}
M.~Lothaire.
\newblock {\em Combinatorics on Words}.
\newblock Cambridge University Press, 1997.

\bibitem{lot:alg:unPat}
M.~Lothaire.
\newblock {\em Algebraic Combinatorics on Words}, chapter~3.
\newblock Cambridge University Press, Cambridge, New York, 2002.

\bibitem{lot:alg}
M.~Lothaire.
\newblock {\em Algebraic Combinatorics on Words}.
\newblock Cambridge University Press, Cambridge, New York, 2002.

\bibitem{Lyndon60}
R.~C. Lyndon.
\newblock Equations in free groups.
\newblock {\em Transactions of the American Mathematical Society}, 96:445--457,
  1960.

\bibitem{Lyndon77}
R.~C. Lyndon and P.~E. Schupp.
\newblock {\em Combinatorial Group Theory}.
\newblock Springer, 1977.

\bibitem{mak:the}
G.~S. {M}akanin.
\newblock The problem of solvability of equations in a free semigroup.
\newblock {\em Matematicheskii Sbornik}, 103:147--236, 1977.

\bibitem{ManSchNow16}
F.~Manea, D.~Nowotka, and M.~L. Schmid.
\newblock On the solvability problem for restricted classes of word equations.
\newblock In {\em Proc. 20th International Conference on Developments in
  Language Theory, {DLT 2016}}, volume 9840 of {\em Lecture Notes in Computer
  Science}, pages 306--318. Springer, 2016.

\bibitem{mat:fin}
Alexandru Mateescu and Arto Salomaa.
\newblock Finite degrees of ambiguity in pattern languages.
\newblock {\em RAIRO Informatique Th\'eoretique et Applications}, 28:233--253,
  1994.

\bibitem{MateescuSalomaa1997}
Alexandru Mateescu and Arto Salomaa.
\newblock Aspects of classical language theory.
\newblock In {\em Handbook of Formal Languages {(1)}}, pages 175--251. 1997.

\bibitem{ng:dev}
Yen~K. Ng and Takeshi Shinohara.
\newblock Developments from enquiries into the learnability of the pattern
  languages from positive data.
\newblock {\em Theoretical Computer Science}, 397:150--165, 2008.

\bibitem{ord:apaJournal}
Sebastian Ordyniak and Alexandru Popa.
\newblock A parameterized study of maximum generalized pattern matching
  problems.
\newblock {\em Algorithmica}, 75:1--26, 2016.

\bibitem{Petit2011}
Jordi Petit.
\newblock Addenda to the survey of layout problems.
\newblock {\em Bulletin of the {EATCS}}, 105:177--201, 2011.
\newblock URL: \url{http://eatcs.org/beatcs/index.php/beatcs/article/view/98}.

\bibitem{Pla2006}
W.~Plandowski.
\newblock An efficient algorithm for solving word equations.
\newblock In {\em Proceedings of the 38th Annual {ACM} Symposium on Theory of
  Computing, STOC 2006}, pages 467--476, 2006.

\bibitem{Reidenbach2006}
Daniel Reidenbach.
\newblock A non-learnable class of e-pattern languages.
\newblock {\em Theor. Comput. Sci.}, 350(1):91--102, 2006.

\bibitem{Reidenbach2007}
Daniel Reidenbach.
\newblock An examination of ohlebusch and ukkonen's conjecture on the
  equivalence problem for e-pattern languages.
\newblock {\em Journal of Automata, Languages and Combinatorics},
  12(3):407--426, 2007.

\bibitem{rei:dis}
Daniel Reidenbach.
\newblock Discontinuities in pattern inference.
\newblock {\em Theoretical Computer Science}, 397:166--193, 2008.

\bibitem{rei:patIaC}
Daniel Reidenbach and Markus~L. Schmid.
\newblock Patterns with bounded treewidth.
\newblock {\em Information and Computation}, 239:87--99, 2014.

\bibitem{Schmid2013}
Markus~L. Schmid.
\newblock A note on the complexity of matching patterns with variables.
\newblock {\em Inf. Process. Lett.}, 113(19-21):729--733, 2013.

\bibitem{sch:wor}
K.~U. Schulz.
\newblock Word unification and transformation of generalized equations.
\newblock {\em Journal of Automated Reasoning}, 11:149--184, 1995.

\bibitem{shi:pol2}
Takeshi Shinohara.
\newblock Polynomial time inference of pattern languages and its application.
\newblock In {\em Proceedings of 7th IBM Symposium on Mathematical Foundations
  of Computer Science, MFCS}, pages 191--209, 1982.

\bibitem{ThilikosSB05}
Dimitrios~M. Thilikos, Maria~J. Serna, and Hans~L. Bodlaender.
\newblock Cutwidth {I:} {A} linear time fixed parameter algorithm.
\newblock {\em J. Algorithms}, 56(1):1--24, 2005.
\newblock \href {http://dx.doi.org/10.1016/j.jalgor.2004.12.001}
  {\path{doi:10.1016/j.jalgor.2004.12.001}}.

\bibitem{zheng2017}
Yunhui Zheng, Vijay Ganesh, Sanu Subramanian, Omer Tripp, Murphy Berzish,
  Julian Dolby, and Xiangyu Zhang.
\newblock Z3str2: an efficient solver for strings, regular expressions, and
  length constraints.
\newblock {\em Formal Methods in System Design}, 50(2-3):249--288, 2017.

\end{thebibliography}

\end{document}